\input amstex
\magnification=1200
\documentstyle{amsppt}
\TagsOnRight
\NoRunningHeads
\NoBlackBoxes

\baselineskip 15pt

\hsize 6.25truein
\vsize = 8.75truein


\catcode`\@=11
\def\logo@{}
\catcode`\@=13

\def\x{\bold x}

\def\det{\text{\rm det}}

\TagsOnRight


\topmatter
\title Symmetries of Higher Dimensional Black Holes\endtitle
\author Vincent Moncrief \\
Department of Mathematics\\ Department of Physics\\
Yale University, New Haven, CT  06520 \\
vincent.moncrief\@yale.edu \\
\medskip
James Isenberg \\
Department of Mathematics \\
University of Oregon \\ 
Eugene, OR 97403 \\
isenberg\@uoregon.edu \endauthor

\abstract We prove that if a stationary, real analytic, asymptotically flat vacuum black hole spacetime of dimension $n\geq 4$ contains a non-degenerate horizon with compact cross sections that are transverse to the stationarity generating Killing vector field then, for each connected component of the black hole's horizon, there is a Killing field which is tangent to the generators of the horizon.  For the case of rotating black holes, the stationarity generating Killing field is not tangent to the horizon generators and therefore the isometry group of the spacetime is at least two dimensional.  Our proof relies on significant extensions of our earlier work on the symmetries of spacetimes containing a compact Cauchy horizon, allowing now for non closed generators of the horizon.
\endabstract
\endtopmatter

\head I.  Introduction\endhead
\vskip .10in

Known examples of black holes [1, 2] and black rings [3, 4] in higher dimensional Einsteinian gravity assure the existence of a rich variety of such objects whose mathematical properties are only just beginning to be uncovered. Whereas the physical motivation for studying general relativity in higher dimensions remains somewhat speculative, it is nevertheless certainly of interest mathematically to ask how far the known properties of black holes in four dimensions extend to hold for black objects in higher dimensions.  Open problems in this area concern the richness of the parameter space for such objects and the issue of their dynamical stability or lack thereof [5, 6]. Furthermore, one would like to know to what extent the laws of black hole thermodynamics can be extended to apply to black objects in higher dimensions.

With respect to this last issue a fundamental question is whether a higher dimensional black object, in the stationary case, necessarily admits a well defined, constant, surface gravity (and corresponding Hawking temperature) associated to its event horizon.  More precisely, one would like to know whether each such object necessarily admits a Killing field $Y = Y^\alpha \frac{\partial}{\partial x^\alpha}$, tangent to the null generators of its horizon and satisfying, on that hypersurface $\Cal H$, an equation of the form
$$
(Y^\beta \nabla_\beta Y^\alpha - \kappa Y^\alpha)\mid_{\Cal H} = 0  \tag1.1
$$
for a suitable constant $\kappa\geq 0$. This constant, when $Y$ is appropriately normalized, would be the surface gravity one is seeking to define for such a black object and the proof of its existence would provide a fundamental component for the generalized laws of black hole thermodynamics that one hopes to formulate in higher dimensions [7, 8].

In this paper, we prove that every stationary, real analytic, asymptotically flat, vacuum spacetime incorporating a non-degenerate horizon with compact  cross sections necessarily admits, for each connected component of its horizon, a Killing field $Y$, tangent to the null generators of that component and satisfying equation (1.1) thereon.  Furthermore, we prove that if the horizon generating Killing field $Y$ of a particular connected component of the horizon fails to coincide (even after rescaling) with the Killing field $T$ generating the one-parameter group of stationary isometries of the spacetime, then there exists one or more additional spacetime Killing fields $\{ \Phi_i \mid i = 1,\ldots,\ell\geq 1\}$, of rotational type, and corresponding (angular velocity) constants $\{ \Omega_i \mid i = 1,\ldots\ell\}$ such that
$$
Y = T + \Sigma^\ell_{i=1} \Omega_i \Phi_i \tag1.2
$$
with 
$$\align
[Y,T] & = [Y,\Phi_i] = [T_, \Phi_i] \tag1.3 \\
& = [\Phi_i, \Phi_j] = 0.
\endalign
$$
Note that if the horizon has several connected components then the constants $\{\Omega_i\}$ and $\kappa$ may vary from one component to another so that, whereas $T$ is fixed once and for all, the particular Killing field $Y$ generating a given component of the horizon may vary with that component.

The qualification `non-degenerate', as used above, means that each horizon generating null geodesic must be incomplete in one temporal direction and corresponds to the requirement that the surface gravity $\kappa$ must be non-vanishing.  Degenerate horizons with $\kappa = 0$ do exist of course and correspond to extremal black holes but our methods are currently incapable of treating them except in special cases which we shall elucidate later.

Asymptotic flatness plays a rather peripheral role in our analysis but insures that one can always find a cross section for each connected component of the horizon to which $T$ is transversal (and in particular non-vanishing), a feature that we shall need to exploit, and furthermore that once the horizon generating Killing field $Y$ is produced, one can normalize it appropriately to yield a well-defined surface gravity $\kappa$ for that component.  Our methods however, could certainly be applied to more general problems in which, for example, a breakdown of asymptotic flatness could be interpreted to represent the effects of material sources `at infinity' on the spacetime geometry.  We also anticipate that the explicit inclusion of certain special types of material sources (such as scalar or electromagnetic fields) into the system would offer no significant extra difficulty but, to keep the analysis as simple as possible, we shall not deal with such enlarged sets of field equations, nor with the inclusion of a cosmological constant here but instead restrict our attention to the pure vacuum case.

Our work on the present problem developed out of ongoing, but somewhat related, research on the question of whether compact Cauchy horizons admitted by real analytic (electro-) vacuum solutions to the $4$-dimensional Einstein equations, must necessarily be Killing horizons.  We answered this question affirmatively for those special cases in which the horizon generating null geodesics were all assumed to be closed curves [9, 10] and we also showed that if a horizon generating Killing field (analogous to $Y$ in the discussion above) exists in cases for which the null generators are not all closed then one or more additional Killing fields must also exist which, together with $Y$, generate a certain (commutative) toral action of isometries of the spacetime [11].  The case of closed generators for the cosmological problem corresponds, at least roughly, to that for which $T$ and $Y$ coincide in the black hole problem and no additional Killing symmetries are implied by the argument (though they could still arise `accidentally' as in the case of static, spherically symmetric black holes).

Our approach here is to map the black hole problem to a corresponding `cosmological' one by translating a slice through the black hole's horizon by a fixed lapse, $s^* > 0$, along the flow generated by the stationary Killing field $T$ and then identifying the (necessarily isometric) `initial' and `final' slices to get a compact horizon of the cosmological type. In $3+1$ dimensions, when the horizon's cross sections are diffeomorphic to $\bold S^2$, one can always choose $s^*$ in such a way that the resultant compact horizon generators are all closed and then appeal to results in [9] (in the analytic case) to prove the existence of the  horizon generating Killing field $Y$ which can then be lifted back up to the original (covering) spacetime [12, 13].  In higher dimensions this simplification is not always possible but, in those special instances when compactification with closed generators is possible, there is no essential difficulty in extending the results of [9] (including those for the degenerate case with $\kappa = 0$) to the higher dimensional problem.  Here though we want to deal with the case of `non-closeable' generators for which a corresponding cosmological result is not yet available.  

The reason why the black hole problem is tractable, even in higher dimensions, while the original four-dimensional cosmological problem remains open, is traceable to the extra structure inherent, a priori, in the former.  This extra structure arises primarily from the presence of the stationary Killing field $T$ and from the fact that the compactification procedure outlined above yields not only a compact horizon of simple topology (namely a trivial circle bundle over a compact manifold) but also a very convenient analytic foliation of this horizon (generated by the flow of the `initial' slice as described above).  By contrast, in the cosmological problem nothing is given a priori other than compactness of the horizon, analyticity of the metric and satisfaction of the field equations.  Any additional structure must then be deduced from this rather meager input information and in particular the horizon may admit no global cross sections whatsoever (the relevant circle bundles being non-trivial for example).  Nevertheless, the $4$-dimensional cosmological problem is, to a large extent, tractable and we anticipate completing our research in this area in the near future.  By contrast,  because of the extra topological complexities alluded to above, the cosmological problem in higher dimensions is largely unexplored territory.

Our work on the higher dimensional black hole problem was stimulated by discussions with A. Ishibashi and S. Hollands who, together with R. Wald, have proven a very similar result to the one reported here [13].  To some extent they avoid the (artificial) analyticity assumption that we impose from the outset but can thereby only deduce the existence of additional Killing fields in the black hole's interior [14].  They must then reinstate the analyticity hypothesis in order to propagate these extra Killing fields to the exterior regions.  Furthermore, they focus upon first producing one or more of the rotational Killing fields ($\{\Phi_i\mid i=1,\ldots, \ell\geq 1\}$ in our notation) whereas we concentrate on producing the horizon generating field $Y$ and then deducing the existence (when $Y\neq T$) of the rotational Killing fields as a corollary to our previous work [11].  

The main technical difficulty in our approach is that, whereas we can easily define a candidate for $Y\mid_{\Cal H}$, the restriction of the hypothetical horizon generating Killing field $Y$ to $\Cal H$, we cannot propagate it off the horizon using our techniques unless we can prove that $Y\mid_{\Cal H}$ is analytic.  We do this by systematically `thickening' the horizon by a complexification procedure that leads to a so-called `Grauert tube' over $\Cal H$, a complex manifold containing the original $\Cal H$ as its real section.  We then extend various real analytic geometric fields defined on $\Cal H$ to holomorphic ones defined on an appropriately chosen Grauert tube and exploit a theorem guaranteeing that holomorphic fields over suitable domains form a Banach space with respect to the $C^0$ norm of uniform convergence, a fact that would not be true in the purely real analytic setting.  Within this setup, we then prove that a sequence of holomorphic approximations to our candidate vector field forms a Cauchy sequence that hence converges to a holomorphic vector field whose real analytic section is the initial data, $Y\mid_{\Cal H}$, that we need.  A Cauchy Kowalewski argument, which makes heavy use of the Einstein equations in the real analytic setting, suffices to propagate $Y\mid_{\Cal H}$ off of $\Cal H$ into the embedding spacetime and to show that the propagated field $Y$ is indeed a spacetime Killing field.  An advantage to our approach is that we provide an explicit formula (rather than just an existence theorem) for the vector field $Y\mid_{\Cal H}$.  We also side-step the need to invoke the machinery of von Neumann's ergodic theorem which played an essential role in the argument of Hollands, et.al.

While analyticity seems to be an artificial restriction to impose a priori on our spacetimes, it is nevertheless unlikely that the results derived here or in \cite{13} can be significantly generalized in scope to remove the analyticity condition completely.  First of all, exterior to any ergo-regions surrounding the black holes in question, the stationarity generator $T$ will be timelike.  On the other hand, the Killing horizon generator $Y$, the existence of which the arguments are intended to establish for each connected component of the horizon, though null on the horizon itself, is anticipated to be timelike in an open region immediately exterior to the horizon.  When Einstein's field equations are reduced with respect to a timelike Killing field, they lead directly to an elliptic system for the metric functions to which Morrey's theorem applies \cite{15}.  The main consequence of this theorem for the present problem is that the solutions are then {\it necessarily analytic}.  In other words, the metric should be automatically analytic wherever it is locally stationary. On the other hand, while the interiors of black holes are expected to be `dynamical' (and thus not amenable to an application of Morrey's theorem), the aforementioned argument of Hollands, et.al. does not require analyticity for the extension of horizon generating Killing fields into the interior regions.

We organize this paper as follows:  In Section II we describe in more detail the hypotheses which characterize the black hole spacetimes that we shall study and state our main theorem.  In Section III we carry out some of the preliminary geometric steps in the proof.  In particular, we a) formulate the identifications that project a stationary black hole spacetime to a corresponding `cosmological' quotient and define a convenient foliation of its compactified horizon, b) set up so-called gaussian null coordinates on a tubular neighborhood of the compactified horizon and recall the form of Einstein's field equations in gaussian null coordinates, c) sketch the proof that the Riemannian metric induced on the leaves of the foliated horizon by the spacetime metric is invariant with respect to flow along the horizon's generators, d) apply Poincar\'e recurrence along the flow of the non closed generators of the compactified horizon to show that the generators `almost  close' in a well-defined sense, and e) set up suitable `ribbon arguments' to prove key estimates that will control the evolution of geometric quantities  along neighboring horizon generators.  Next in Section IV we discuss the consequences of the non-degeneracy condition and prove, using a ribbon argument, that either all the generators of a connected horizon are incomplete in one direction (the non-degenerate case) or else that they are all complete in both directions (the degenerate case).  We then proceed in Section V to define a vector field on the horizon which serves as our candidate for the boundary data of our ultimate horizon generating Killing field.  In Section VI, we prove the analyticity of our candidate vector field and in Section VII show how to extend this candidate to an analytic vector field on a tubular neighborhood of the horizon and to prove that it satisfies Killing's equations thereon.  We also show that the new Killing field commutes with the stationarity generator $T$.  In Section VIII, we discuss the lifting of our horizon generating Killing field back to a tubular neighborhood of the non compact horizon and its further analytic extension to the originally given black hole spacetime.  This same section discusses the full black hole isometry group and, in particular, the origin of the `axisymmetry generators' $\{\Phi_i \mid i=1,\ldots, \ell\}$ and corresponding angular velocities $\{\Omega_i \mid i = 1,\ldots, \ell\}$.  The role of `topological censorship' in ensuring that the aforementioned Killing fields can be analytically extended to the full {\it domain of outer communications} of the given black hole spacetime is also discussed in Section VIII.  Finally, in Section IX we discuss the extent to which our techniques can be applied to degenerate horizons.

\newpage

\head II. Statement of our main theorem\endhead
\vskip .10in

By definition a spacetime $(^{(n+1)}\tilde V, \tilde g)$  contains a black hole if it is asymptotically flat in the sense of admitting a conformal asymptotic structure whose boundary includes a future null infinity $\Cal I^+$ and if the complement with respect to $^{(n+1)}\tilde V$ of the causal past, $J^-(\Cal I^+)$, of $\Cal I^+$ is
nonempty\footnote"$^{1}$"{We place a tilde $(\sim )$ over the black hole spacetime structures (such as the manifold  $^{(n+1)}\tilde V$, metric $\tilde g$, stationarity generator $\tilde T$, etc.) to distinguish these from the corresponding structures for the spatially compactified, ``cosmological'' spacetime introduced below.  This choice is governed by the fact that most of our analysis is carried out in the cosmological setting and thus leads to a simplification of the notation throughout the bulk of the article}.  This complement, $^{(n+1)}\tilde V\backslash J^{-}(\Cal I^+)$, is called the black hole region of the spacetime and its boundary, $\partial(^{(n+1)}\tilde V\backslash J^-(\Cal I^+))$, is called the future event horizon. If the conformal asymptotic structure also includes a past null infinity $\Cal I^-$ and if the complement of the causal future, $J^+(\Cal I^-)$, of $\Cal I^-$ is nonempty, we call $^{(n+1)}\tilde V\backslash J^+(\Cal I^-)$ the {\it white hole} region of spacetime and its boundary, $\partial(^{(n+1)}\tilde V\backslash J^+(\Cal I^-))$, the corresponding past horizon.  As in well-known examples the past and future horizons could intersect in a `bifurcation surface' 
$$
\Cal S := \partial (^{(n+1)}\tilde V\backslash J^-(\Cal I^+))\cap\partial (^{(n+1)}\tilde V\backslash J^+(\Cal I^-)).
$$
Let $\tilde \Cal H$ designate a connected component of $\partial (^{(n+1)}\tilde V\backslash J^-(\Cal I^+))\backslash\Cal S$.  For brevity, we shall refer to $\tilde\Cal H$ as the future horizon (even though $\partial (^{(n+1)}\tilde V\backslash J^-(\Cal I^+))\backslash\Cal S$ could involve the disjoint union of several such components).
We shall assume that $\tilde\Cal H$ has the topology $\Sigma\times\bold R$ where $\Sigma$ is a compact (connected)
$(n-1)$-dimensional manifold.  If the event horizon contains more than one connected component, the manifold $\Sigma$ could vary from one component to the next. Although we shall formulate our arguments specifically for future horizons, they would apply equally well to past horizons --- the latter defined as connected components of $\partial (^{(n+1)}\tilde V\backslash J^+(\Cal I^-))\backslash\Cal S$.

The spacetimes for which our results hold must satisfy four additional conditions beyond the requirement that they contain a black hole with the topological property described above.  First, we require that the spacetime manifold $^{(n+1)}\tilde V$, the spacetime metric $\tilde g$ and the embedding of $\tilde\Cal H$ in $^{(n+1)}\tilde V$ all be real analytic and that the metric satisfy the vacuum Einstein equations $G_{\mu\nu}(\tilde g) = 0$.  Next, we require that the null geodesic generators of the horizon be geodesically incomplete in one direction -  the `past' direction.  Strictly speaking, we need only require this of a single generator in each connected component $\tilde\Cal H$ of the horizon since, as we shall show in Section IV, if a single generator of $\tilde\Cal H$ is incomplete to the past, then all must be and furthermore every generator of $\tilde\Cal H$ is then in fact complete to the future.  This condition, which we shall label ``nondegeneracy'' of the horizon, plays a crucial role in our construction of the candidate for the horizon generating Killing field and we currently do not know how to remove this hypothesis except in the exceptional, special case of what we shall call ``closeable'' generators.  On the other hand, degenerate black hole solutions with non-closeable generators do exist so this leaves an important gap to be filled.  We note also that, in the nondegenerate case of most interest here, the ideal past endpoints of incomplete generators of $\tilde\Cal H$ define a certain boundary thereto (the ``bifurcation surface'')  which plays a key role in the arguments of Hollands, et. al. to partially eliminate the analyticity requirement. As discussed in their article however, these arguments only succeed to allow one to propagate the horizon generating Killing fields to the interior regions of the black holes in question and not to their exteriors.  For propagation to the exterior regions, they require analyticity as we do and exploit the Cauchy-Kowalewski theorem, and a theorem of Nomizu's discussed further below, to complete the construction.  On the  other hand, as discussed in the Introduction, it seems plausible that stationary vacuum black hole exteriors must necessarily be analytic in which case the hypothesis of analyticity is not as artificial as it at first seems to be.

Our third requirement is that the spacetime $(^{(n+1)}\tilde V, \tilde g)$ admit a Killing field $\tilde T$ which is asymptotically (in a neighborhood of infinity) timelike and which has complete orbits.  Such a spacetime is called ``stationary'' and we shall refer to $\tilde T$ as its ``stationarity generator'' since it generates a one-parameter group of isometries of  $(^{(n+1)}\tilde V, \tilde g)$ for which the action is asymptotically timelike.  Since the action generated by $\tilde T$ preserves any natural geometric structure defined for $(^{(n+1)}\tilde V, \tilde g)$, $\tilde T$ must be tangent to $\tilde\Cal H$.  In the special case that $\tilde T$ happens to be tangent to the null generators of every connected component of the horizon, then of course $\tilde T$ would be null on each connected component $\tilde\Cal H$, hence normal to this hypersurface, and our arguments give no further information about the structure of $(^{(n+1)}\tilde V, \tilde g)$.  By definition, this is the case of ``non-rotating'' black holes whereas our arguments are all formulated to deal with ``rotating'' black holes for which (by definition) $\tilde T$ is not normal to $\tilde\Cal H$.  Henceforth,  we restrict our attention to the rotating case.

The fourth and final restriction we impose on our black hole spacetimes, in addition to the aforementioned topological condition that each $\tilde\Cal H$ be diffeomorphic to $\Sigma\times\bold R$, for some compact $(n-1)$-manifold $\Sigma$, is that there be an analytic embedding $i(\Sigma)$ of $\Sigma$ in $\tilde\Cal H$ such that the Killing field $\tilde T$ is transverse to $i(\Sigma)$.  However, if the spacetime admits a past null infinity $\Cal I^-$ and if the horizon component $\tilde\Cal H$ is contained in the future of $\Cal I^-$, then (recalling that $\tilde T$ is timelike near infinity, hence near $\Cal I^-$ and $\Cal I^+$) it follows from an argument sketched in \cite{7} (see the proof of Proposition 4.1 in this reference) that we need not independently impose the requirement that $\tilde T$ be transverse to a suitably chosen $i(\Sigma)$ - the existence of such an embedding is automatic.  Since however one may not always wish to assume this additional property of the asymptotic structure we explicitly include the transversal embedding requirement in our hypotheses.

We now state our main result:

\proclaim{Theorem 2.1} Let $(^{(n+1)}\tilde V, \tilde g)$ be an analytic, asymptotically flat, vacuum black hole spacetime (with stationarity generator $\tilde T$ that is timelike near infinity) and for which at least one of the associated black holes is rotating and non-degenerate.  Assume that the corresponding connected component, $\tilde\Cal H$, of the horizon is diffeomorphic to $\Sigma\times\bold R$ for some compact $(n-1)$-manifold $\Sigma$ and that $\tilde\Cal H$ is analytically embedded in $(^{(n+1)}\tilde V, \tilde g)$ and that there is an analytic embedding $i(\Sigma)$ of $\Sigma$ in $\tilde \Cal H$ such that $i(\Sigma)$ is transverse to the stationarity generator $\tilde T$. Then the spacetime admits an independent Killing field $\tilde Y$ which is tangent to the null generators of $\tilde\Cal H$ and $\ell\geq 1$ additional Killing fields $\{\Phi_i\mid i=1,\ldots, \ell\}$ that are spacelike on $\tilde\Cal H$, generate an $\ell$-dimensional toral action on spacetime and satisfy
$$
\tilde Y = \tilde T + \sum\limits^{\ell}_{i=1} \Omega_i \tilde\Phi_i
$$
for suitable (angular velocity) constants $\{ \Omega_i\mid i = 1,\ldots , \ell\}$.  These Killing fields satisfy
$$\align
&[\tilde T, \tilde Y] = [\tilde Y, \tilde\Phi_i] = [\tilde T, \tilde\Phi_i] \\
&= [\tilde\Phi_i, \tilde\Phi_j] = 0 \,\, \text{and} \\
&(Y^\beta \tilde\nabla_\beta Y^\alpha - \kappa Y^\alpha)_{\tilde\Cal H} = 0
\endalign
$$
for a suitable constant $\kappa > 0$ (where $\tilde\nabla_\beta$ signifies covariant differentiation with respect to $\tilde g$).
\endproclaim

As already mentioned in the introduction, if the horizon admits several (rotating, non-degenerate) components, then the constants $\Omega_i$ and $\kappa$ could vary from one component to another so that, whereas $\tilde T$ is fixed once and for all, the Killing generator $\tilde Y$ which is null on a given component $\tilde\Cal H$ may vary with the choice of that component.  That the constant $\kappa$ is non-vanishing for a given component $\tilde\Cal H$, results from the non-degeneracy of that component.  Degenerate black hole solutions do of course exist, but except for certain special cases wherein the associated (rotating) degenerate horizons admit ``closeable generators'' (in a sense that we shall define below) our techniques are not currently applicable to them.

\head III.  Basic Analytic tools and geometric constructions\endhead
\vskip .10in

In this section, we begin by introducing an analytic foliation of $\tilde\Cal H$ defined by letting the chosen cross-section, $i(\Sigma)$, flow along the integral curves of the stationarity generator $\tilde T$. Relative to this foliation we decompose $\tilde T\mid_{\tilde\Cal H}$ into a sum of analytic vector fields, $\tilde X\mid_{\tilde\Cal H}$ and $\tilde S\mid_{\tilde\Cal H}$ where $\tilde X\mid_{\tilde\Cal H}$ is tangent to the null generators of $\tilde\Cal H$ and $\tilde S\mid_{\Cal H}$ is tangent to the leaves of the foliation.  Based on this setup, we then define a system of ``gaussian null'' coordinates on a suitably chosen tubular neighborhood $^{(n+1)}\tilde U$ of $\tilde\Cal H$ in $^{(n+1)}\tilde V$ and then recall the form of $\tilde g$ and its Ricci tensor when these objects are expressed in such coordinates.  We next quotient the tubular neighborhood spacetime, $(^{(n+1)}\tilde U, \tilde g)$, by a discrete subgroup of the $\bold R$-action generated by $\tilde T$ so that the image $\Cal H$, of the horizon component $\tilde\Cal H$ is now compact and diffeomorphic to $\Sigma\times\bold S^1$.  Under the (extremely special) circumstances that the quotient can be chosen so that the null generators of $\Cal H$ are all closed curves, we can apply the methods of Reference \cite{9}, for both degenerate and non-degenerate horizons, to show that $\Cal H$ is in fact a Killing horizon. Since our main interest however, is the ``non-closeable'' case, we need to develop new techniques going beyond those of the foregoing reference. The rest of this section is devoted to introducing some of the needed tools.

\subhead 3.1. Horizon Foliation and Compactification \endsubhead

Let
$$
\tilde\Cal A_s :\,\, ^{(n+1)}\tilde V\to\,\, ^{(n+1)}\tilde V
$$
for $s\in\bold R$ denote the action of the stationarity generator $\tilde T$ on the spacetime manifold $^{(n+1)}\tilde V$. Since $\tilde\Cal A_s$ is an isometric action it preserves the horizon and therefore its generator $\tilde T$ will be automatically tangent to any connected component $\tilde\Cal H$ thereof. We assume however, that $\tilde T$ is not tangent to the null generators of $\tilde\Cal H$ (i.e., that the corresponding black hole is rotating).  As discussed above, we also assume that \newline $\tilde\Cal H \approx\Sigma\times\bold R$ for some compact, connected $(n-1)$ -manifold $\Sigma$, that $\tilde\Cal H$ is analytically embedded in $(^{(n+1)}\tilde V, \tilde g)$ and that there is an analytic embedding, $i(\Sigma)$, of $\Sigma$ in $\tilde\Cal H$ such that $i(\Sigma)$ is everywhere transverse to $\tilde T\mid_{\tilde\Cal H}$ .  We now extend the cross-section $i(\Sigma)$ of $\tilde\Cal H$ to a foliation of the horizon through the action $\tilde\Cal A_s: i(\Sigma)\to\tilde\Cal H$, defining $\Sigma (s):= \tilde\Cal A_s (i(\Sigma))$ as leaves of this foliation for $s\in\bold R$.

We now introduce a coordinate function $x^n$ on $\tilde\Cal H$, constant on each of the aforementioned leaves, by setting $x^n(p)=0$ for $p\in i(\Sigma)$ and imposing $\Cal L_{\tilde T\mid_{\tilde\Cal H}} x^n = 1$ so that  $x^n(q) = s$ for $q\in\Sigma (s)$.  As in Ref. \cite{13}, we next decompose $\tilde T\mid_{\tilde\Cal H}$ by setting 
$$
\tilde T\mid_{\tilde\Cal H} = \tilde X\mid_{\tilde\Cal H} + \tilde S\mid_{\tilde\Cal H} \tag3.1
$$
where $\tilde X\mid_{\tilde\Cal H}$ is tangent to the null generators of $\tilde\Cal H$ and $\tilde S\mid_{\tilde\Cal H}$ is tangent to the leaves of the chosen foliation.  By virtue of the analyticity of $\tilde g, \tilde\Cal H$ and $\tilde T$ both $\tilde X\mid_{\tilde\Cal H}$ and $\tilde S\mid_{\tilde\Cal H}$ will be analytic vector fields on $\tilde\Cal H$  and, from the fact that $\Cal L_{\tilde T} \tilde g = 0$, it is straightforward to show (by working, for example in coordinates for which  $\tilde T\mid_{\tilde\Cal H} = \frac{\partial}{\partial x^{n}})$ that
$$
[\tilde T\mid_{\tilde\Cal H}, \tilde X\mid_{\tilde\Cal H}] = 0 \tag3.2
$$
and
$$
[\tilde T\mid_{\tilde\Cal H}, \tilde S\mid_{\tilde\Cal H}] = 0. \tag3.3
$$
It follows from the construction that
$$
\Cal L_{\tilde X\mid_{\tilde\Cal H}} x^n = 1,  \quad \Cal L_{\tilde S\mid_{\tilde\Cal H}} x^n = 0 \tag3.4
$$
and thus that the flow on $\tilde\Cal H$ generated by $\tilde X\mid_{\tilde\Cal H}$ takes leaves of the chosen foliation to other such leaves whereas the flow generated by $\tilde S\mid_{\tilde\Cal H}$ leaves each leaf invariant.

We now introduce complementary `spatial' coordinates $\{ x^a \mid a=1,\ldots, n-1\}$ on $\tilde\Cal H$ by choosing them arbitrarily on $i(\Sigma)$ and ``dragging'' them along the integral curves of $\tilde X\mid_{\tilde\Cal H}$ (i.e., imposing $\Cal L_{\tilde X\mid_{\tilde\Cal H}} x^a = 0$ so that the new coordinates are constant along the null generators of $\tilde\Cal H$).  Several sets of such coordinates are of course needed to cover the compact manifold $i(\Sigma)$, and thus label all the null generators of $\tilde\Cal H$, but our subsequent constructions will all be naturally covariant with respect to transformations among these different charts.

With the foregoing constructions in place, there exists a unique, analytic, null vector field $\tilde L\mid_{\tilde\Cal H}$ defined on $\tilde\Cal H$ (and everywhere transverse to this horizon) by the requirements that
$$\align
&\tilde g_{\alpha\beta} \tilde L^\alpha\mid_{\tilde\Cal H} \tilde L^\beta\mid_{\tilde\Cal H} = 0, \tag3.5 \\
&\tilde g_{\alpha\beta} \tilde L^\alpha\mid_{\tilde\Cal H} \tilde X^\beta\mid_{\tilde\Cal H} = 1
\endalign
$$
and that $\tilde L\mid_{\tilde\Cal H}$ be orthogonal to the leaves $\Sigma (s)$ of the chosen foliation.  The null geodesics of $(^{(n+1)}\tilde V, \tilde g)$ determined by $\tilde L\mid_{\tilde\Cal H}$ will be non-intersecting on a  suitably chosen tubular neighborhood $^{(n+1)}\tilde U\approx \tilde\Cal H \times (-a,a)$ of $\tilde\Cal H$ in $^{(n+1)}\tilde V$ and, from the invariance of $\tilde g$ with respect  to the action generated by $\tilde T$, one can always choose this tubular neighborhood so that it consists entirely of complete orbits of $\tilde T$.  Now define a coordinate function $t$ on $^{(n+1)}\tilde U$ by setting $t=0$ on $\tilde\Cal H$ and taking  $t(p)$ to equal the value of the affine parameter of the null geodesic determined by $\tilde L\mid_{\tilde\Cal H}$ that ``starts'' on $\tilde\Cal H$ (with parameter initially zero) and arrives at $p$ (before exiting $^{(n+1)}\tilde U$).  Next, extend the coordinate functions $\{ (x^a, x^n)\mid a=1,\ldots ,n-1\}$ to coordinates on $^{(n+1)}\tilde U$ by holding each function  constant along the aforementioned null geodesics and define the tangent field $\tilde L$ to these geodesics by setting $\tilde L = \frac{\partial}{\partial t}$ on $^{(n+1)}\tilde U$.  Clearly, $\tilde L$ coincides with $\tilde L\mid_{\tilde\Cal H}$ at points of the horizon.  Similarly, the vector field $\tilde X$ defined on $^{(n+1)}\tilde U$ by $\tilde X = \frac{\partial}{\partial x^n}$, clearly coincides by construction with the vector field $\tilde X\mid_{\tilde\Cal H}$ at points of $\tilde\Cal H$.

Coordinates of this type were introduced for convenience in Reference \cite{9} and referred to there as {\it gaussian null coordinates} because of their resemblance to ordinary gaussian normal coordinates (based on timelike geodesics normal to a spacelike hypersurface).  The analyticity of the construction of such coordinates was proved in \cite{9} (by an argument which is easily seen to be independent of spacetime dimension) and the metric form of $\tilde g$ (restricted to $^{(n+1)}\tilde U$) was therein found to be
$$\align
\tilde g &= \tilde g_{\mu\nu} dx^\mu\otimes dx^\nu \tag3.6 \\
&= dt\otimes dx^n + dx^n\otimes dt + \tilde\varphi dx^n\otimes dx^n \\
&+ \tilde\beta_a dx^a\otimes dx^n + \tilde\beta_a dx^n \otimes dx^a \\
&+ \tilde\mu_{ab} dx^a\otimes dx^b 
\endalign
$$
where $\tilde\varphi, \tilde\beta_a dx^a$ and $\tilde\mu_{ab} dx^a\otimes dx^b$ may be viewed (respectively, for fixed values of $(t, x^n)$) as a scalar, one-form and $(n-1)$-dimensional Riemannian metric with respect to transformations of the form $x^{a'} = f^a (x^1,\ldots, x^{n-1})$.  By virtue of the construction of these coordinates, both $\tilde\varphi$ and $\tilde\beta_adx^a$ vanish on the hypersurface $\tilde\Cal H$ (i.e. for $t=0$) and $\tilde\mu_{ab} dx^a\otimes dx^b$ induces a Riemannian structure  on each of the leaves $\Sigma (s)$ of $\tilde\Cal H$.

Einstein's field equations for the gaussian null metric form were written out explicitly in Equations (2.9) of Reference \cite{9} for the case of $4$-dimensional spacetimes.  These formulas however, apply equally well to $(n+1)$-dimensions provided that we extend the range of indices $a,b,c, \ldots$ from $\{ 1,2\}$ to $\{ 1,\ldots ,n-1\}$, replace $\frac{\partial}{\partial x^3}$ by $\frac{\partial}{\partial x^n}$,  $^{(2)}\nabla_a$ by $^{(n-1)}\tilde\nabla_a$ and $^{(2)}R_{ab}$ by $^{(n-1)}\tilde R_{ab}$ where $^{(n-1)}\tilde\nabla_a$ represents covariant differentiation with respect to the metric $\tilde\mu_{ab} dx^a\otimes dx^b$ and $^{(n-1)}\tilde R_{ab}$ is the Ricci tensor of this metric.  The proof of existence of a spacetime Killing field tangent to the null generators of $\tilde\Cal H$ will make heavy use of the specific form of these field equations and, beyond a certain point, will parallel the arguments given in Reference \cite{9} for the case of closed (null geodesic) generators.

Unlike ordinary gaussian normal coordinates, gaussian null coordinates depend upon the choice of a foliation $\Sigma (s)$ of the ``initial'' hypersurface $\tilde\Cal H$ and a vector field $\tilde X\mid_{\tilde\Cal H}$ tangent to the null generators of that hypersurface (whose flow preserves the leaves of the given foliation).  In the present context, when we wish to emphasize that the coordinates have been adapted to the choice of such a foliation of $\tilde\Cal H$ and null tangent field thereon $\tilde X\mid_{\tilde\Cal H}$, we shall refer to them as {\it adapted gaussian null} coordinates or {\it agn} coordinates for brevity.

By writing out the Killing equations for $\tilde T$ in the coordinate system defined above, it is straightforward to show that $\tilde T$ takes the form
$$
\tilde T = \frac{\partial}{\partial x^n} + \tilde T^b (x^1,\ldots, x^{n-1})\frac{\partial}{\partial x^b} \tag3.7
$$
and that the metric functions $\{\tilde\varphi, \tilde\beta _a, \tilde\mu_{ab}\}$ satisfy
$$\align
\tilde\varphi_{,n} &= - \tilde T^a \tilde\varphi_{,a} \tag3.8 \\
\tilde\beta_{a,n} &= - (\tilde T^b \tilde\beta_{a,b} + \tilde T^b_{,a} \tilde\beta_b) \\
\tilde\mu_{ab,n} &= - (\tilde T^c \tilde\mu_{ab,c} + \tilde T^c_{,a} \tilde\mu_{cb} + \tilde T^c_{,b} \tilde\mu_{ac}) 
\endalign
$$
on $^{(n+1)}\tilde U$.  Thus, in particular, one finds that
$$
[\tilde T , \frac{\partial}{\partial x^n}] = [\tilde T , \frac{\partial}{\partial t} ] = 0 \tag3.9
$$
and that
$$
\tilde T x^n = 1 \tag3.10
$$
on $^{(n+1)}\tilde U$.   It follows from (3.10) that the leaves, $x^n =$ constant, of the foliation of $^{(n+1)}\tilde U$ defined by our coordinate construction, are dragged into one another by the flow generated by $\tilde T$.

We now wish to quotient $(^{(n+1)}\tilde U, \tilde g)$ by the action of a discrete subgroup of the isometry group generated by $\tilde T$ so as to compactify the horizon.  Choosing $s^* > 0$, let $\tilde\Cal A_{s^*}$ generate the (nontrivial) isometry for this discrete group action and designate the corresponding quotients of $\tilde\Cal H$ and $^{(n+1)}\tilde U$ by $\Cal H$ and $^{(n+1)}U$ respectively.  Since $\tilde g$ is invariant with respect to this group action, it naturally induces a metric on the quotient manifold which we shall designate by $g$.  The pair $(^{(n+1)}U, g)$ defines a certain vacuum, `cosmological' spacetime that includes an embedded compact null hypersurface $\Cal H$ that is diffeomorphic, by construction, to $\Sigma\times S^1$.  We shall see later (at least when $\Cal H$ is non-degenerate) that
$(^{(n+1)}U, g)$ is globally hyperbolic on one side of $\Cal H$  but acausal on the other and thus that $\Cal H$ serves as a Cauchy horizon bounding the globally hyperbolic region.  In those (exceptional) cases when $s^*$ can be chosen so that the (null geodesic) generators of $\Cal H$ are all closed curves, one can apply a straightforward generalization of the ($3+1$ dimensional)  arguments given in Reference \cite{9} to prove (even in degenerate cases) the existence of a horizon generating Killing field $Y$ on the cosmological spacetime $(^{(n+1)}U, g)$.  This Killing field can be lifted back to the covering manifold $(^{(n+1)} \tilde U, \tilde g)$ and ultimately (as we shall see) analytically extended to the originally given black hole spacetime $(^{(n+1)}\tilde V,\tilde g)$  or, more precisely stated, at least to the latter's {\it domain of outer communications}.

Since the action of $\tilde\Cal A_{s^*}$ on $^{(n+1)}\tilde U$  maps an $x^n = c =$ constant slice isometrically to the slice $x^n = c + s^*$ while leaving the coordinate function $t$ invariant (c.f., Equation (3.9)), one can represent the manifold $^{(n+1)}U$ concretely by identifying these two slices via a diffeomorphism determined by $\tilde \Cal A_{s^*}$ and expressible (in the chosen coordinates) as
$$
(t, x^1,\ldots, x^{n-1}) \mapsto (t, f^1(x^1,\ldots, x^{n-1}),\ldots, f^{n-1}(x^1,\ldots x^{n-1}).
$$
It is straightforward to verify that the vector fields $\tilde L = \frac{\partial}{\partial t}$ and $\tilde X = \frac{\partial}{\partial x^n}$, defined on $(^{(n+1)}\tilde U, \tilde g)$, pass naturally to the quotient manifold $(^{(n+1)}U, g)$ where they induce vector fields we shall designate by $L$ and $X$ respectively.  We shall refer to coordinates induced on $(^{(n+1)}U, g)$ from agn coordinates defined on $(^{(n+1)}\tilde U, \tilde g)$  also as agn coordinates.  One need only recall that in sweeping through a foliation of $^{(n+1)}U$ the induced agn charts transform according to the formula given above when $x^n$ is mapped to $x^n + s^*$.

For the cases of most interest in this paper, $s^*$ cannot be chosen so that the generators of $\Cal H$ are all closed curves (i.e., the generators of $\tilde\Cal H$ are not `closeable').  To handle these cases, we thus need to develop techniques that go beyond those of Reference \cite{9}.
\medskip

\subhead 3.2. Invariance of the Transversal Metric \endsubhead
\vskip .10in

Consider an arbitrary $(n-1)$ dimensional disk $D$ which is analytically embedded in
$\Cal H$ and which is everywhere transversal to the null generators of
that hypersurface.  In a gaussian null coordinate chart which covers
$D$, it is clear that $D$ has a coordinate characterization of the
form,
$$
t = 0, \quad x^n = f(x^a) \tag3.11
$$
for some real analytic function $f$.  (Here the $\{x^a\}$ range over
those values corresponding to the generators which intercept $D$).
From Equations (3.6), (3.11) and the facts that $\varphi$ and $\beta_a$ vanish on $\Cal H$ one sees that $g$ induces a
Riemannian metric $\mu_D$, given by \footnote"$^{2}$"{We parametrize
$g$ in terms of $\{\varphi, \beta_a dx^a, \mu_{ab}dx^a\otimes dx^b\}$ as in Equation (3.6).}
$$
\mu_D = \mu_{ab}\mid_{t=0, ~~ x^{n}=f(x^{c})} dx^a\otimes dx^b \tag3.12
$$
on $D$.  If we let $D$ flow along the integral curves of the vector
field $X = \frac{\partial}{\partial x^{n}}$ associated (at least
locally) to the chosen chart, then we get a one-parameter family
$D_\lambda$ of embeddings of $D$ in $\Cal H$ characterized by
$$
t = 0, \quad x^n = f(x^a) + \lambda \tag3.13
$$
and a corresponding family of metrics $\mu_{D_{\lambda}}$ given by
$$
\mu_{D_{\lambda}} = \mu_{ab}\mid_{t=0, ~ ~x^{n}=f(x^{c})+\lambda} dx^a\otimes
dx^b. \tag3.14
$$
Here $\lambda$ ranges over some open interval containing $\lambda =
0$. 

Locally one can always choose a particular vector field $K$ tangent
to the null generators of $\Cal H$ such that the integral curves of $K$
coincide with the \it affinely parametrized \rm null geodesics
generating $\Cal H$ (i.e., such that in coordinates locally adapted to $K$ the curves $\{x^\mu(\lambda)\}$
defined by $t(\lambda ) = 0, ~~ x^a(\lambda) = ~\text{constant}$,
$x^n(\lambda ) = \overset{\circ}\to x^n + \lambda$ are affinely
parametrized null geodesics generating (a portion of) $\Cal H$, with
$\lambda$ an affine parameter).  $K$ is of course not unique (since
there is no canonical normalization for $\lambda$ along each
generator) but can be fixed by prescribing it at each point of some
transversal $(n-1)$-manifold.  In general, $K$ may also not be extendable
to a globally defined vector field on $\Cal H$ (since the affinely
parametrized generators of $\Cal H$ may be incomplete whereas the flow of
a globally defined vector field on the \it compact \rm manifold $\Cal H$
must be complete) but this is of no consequence in the following
construction.   For any point $p\in \Cal H$ choose a disk $D$ which
contains $p$ and is everywhere transversal to the null generators of
$\Cal H$.  Construct, on a neighborhood of $D$ in $\Cal H$, a vector field $K$
of the type described above and let $\{x^\alpha\} = \{t, x^n,x^a\}$
be an agn coordinate chart adapted to $K$ (i.e., so that
$\frac{\partial}{\partial x^{n}} = K$ is thus tangent to the affinely
parametrized generators of $\Cal H$).   Now let $D$ flow along the
integral curves of $K$ to get a one-parameter family of embedded
disks $D_\lambda$ and a corresponding family of induced Riemannian
metrics $\mu_{D_{\lambda}}$ as described above.

In terms of this construction, one can compute the \it expansion \rm
$\hat\theta$ of the null generators at $p$ by evaluating
$$
\hat\theta (p) = (\frac{\partial}{\partial\lambda} \ell n\bigl(\det
~\mu_{D_{\lambda}})\bigr)\mid \Sb \lambda =\lambda(p) \\ x^a=x^a(p)
\endSb \tag3.15
$$
It is not difficult to verify that this definition is independent of
the particular choice of transversal manifold $D$ chosen through $p$
and of the particular coordinates $\{x^a\}$ used to label the
generators near $p$.  In fact, this definition of $\hat\theta$ is
equivalent to the usual definition of the \it expansion \rm of the
null generators of a null hypersurface \cite{16}.

In our case, however, $\Cal H$ is not an arbitrary null surface.  It is,
by assumption a compact null surface in a vacuum spacetime.  For
such a hypersurface Hawking and Ellis have proven the important
result that $\hat\theta$ vanishes at every point $p\in\Cal H$ \cite{16}.
Strictly speaking, the original Hawking and Ellis argument was given only for $3+1$ dimensional spacetimes but Hollands, et. al.  have shown in Reference \cite{13} that it generalizes to the $n+1$ dimensional context of primary interest here.
Thus in an agn coordinate chart adapted to $K$ one has
$$
(\det~ \mu_{ab})_{,n}\mid_{t=0} = 0 \tag3.16
$$
at every point of $\Cal H$ covered by the chart.  However, the Einstein
equation $R_{nn} = 0$, restricted to $\Cal H$, yields
$$\align
&\overset{\circ}\to R_{nn} = 0 = [(\ell n \sqrt{\det\mu})_{,nn} \\
&+\frac{1}{2} \varphi_{,t} (\ell n\sqrt{\det\mu})_{,n} \tag3.17 \\
&+ \frac{1}{4} \mu^{ac} \mu^{bd} \mu_{ab,n} \mu_{cd,n}]\mid_{t=0}
\endalign
$$
in an arbitrary gaussian null coordinate chart (where
$(\det\mu)\equiv \det(\mu_{ab})$).  Combining Equations (3.16) and (3.17),
we see that $\mu_{ab,n}\mid_{t=0} = 0$ throughout the local chart
adapted to $K$.

From this result, it follows easily that the metric $\mu_D$ induced
upon an arbitrary disk transversal to a given bundle of null
generators of $\Cal H$ is, in fact, independent of the disk chosen.  To
see this one computes, recalling Equations (3.6) and (3.11), the metric
induced upon an arbitrary such disk $D$ (satisfying $t=0,
x^n=f(x^a)$).  From the result that $\mu_{ab,n}\mid_{t=0} = 0$ it
follows that this induced metric is independent of the function $f$
(which embeds $D$ in the given bundle) and hence of the particular
transversal disk chosen.  Though this calculation was carried out
using a special family of charts, the definition of the induced
metric is a geometrical one and thus the invariance of this metric
(relative to an arbitrary displacement along the null generators of
$\Cal H$) is independent of any choice of charts.

\newpage

\subhead 3.3. Application of the Poincar\'e Recurrence Theorem
\endsubhead
\vskip .10in

In this subsection we shall show that the Poincar\'e recurrence
theorem \cite{17, 18} can be applied to the flow on $\Cal H$ generated by the vector field $X$ defined in Section 3.1.  Using this theorem we shall then show that every point $p\in\Cal H$, when mapped sufficiently
far (in either direction) along the flow of $X$, returns arbitrarily
closely to its initial position. From the constructions defined in Section 3.1, it is easily seen that the vector field
$$\align
V &= (2 + \varphi)^{-\frac{1}{2}} ( L - X) \tag3.18 \\
&= (2 + \varphi)^{-\frac{1}{2}} (\frac{\partial}{\partial t} - \frac{\partial}{\partial x^n}) 
\endalign
$$
is analytic and timelike on some (sufficiently small) tubular neighborhood of the horizon $\Cal H$ (since $\overset{\circ}\to\varphi = \varphi\mid_{\Cal H} = 0$), is transverse to $\Cal H$ and satisfies the normalization condition
$$
g(V, V) = -1. \tag3.19
$$
While this specific vector field is adequate for our purposes here it is convenient, with a view towards possible future applications, to let the symbol $V$ stand below for any vector field defined on a tubular neighborhood of $\Cal H$ that has the corresponding properties of analyticity, transversality to $\Cal H$ and satisfaction of Equation (3.19).

Written out explicitly the normalization condition (3.19) yields
$$
-1 = g(V,V)\mid_{\Cal H} = \{ 2V^tV^n + \mu_{ab} V^aV^b\}\mid_{t=0} \tag3.20
$$
which clearly implies that $V^t\mid_{t=0} = g(X, V)\mid_{\Cal H}$ is nowhere vanishing on $\Cal H$.  Assume for definiteness (as in the example given above) that $V^t\mid_{t=0} > 0$ everywhere on $\Cal H$.  To simplify the discussion, assume that $V$ is defined everywhere on $^{(n+1)}U$ (even though an arbitrarily small tubular neighborhood of $\Cal H$ would suffice for the following).

We now define a positive definite metric $g'$ on $^{(n+1)}U$ by setting
$$
g'(Y,Z) = g(Y,Z) + 2g(Y,V)g(Z,V) \tag3.21
$$
for any pair of vector fields $Y,Z$ defined on $^{(n+1)}U$.  This metric
induces a Riemannian metric $^{(n)}g'$ on $\Cal H$ given, in an arbitrary
gaussian null coordinate chart adapted to $X$, by the expressions
$$\align
^{(n)}g'_{nn} &= (2V^tV^t)\mid_{t=0} \\
^{(n)}g'_{na} &= (2\mu_{ab}V^bV^t)\mid_{t=0} \tag3.22 \\
^{(n)}g'_{ab} &= (\mu_{ab} +2\mu_{ac}V^c\mu_{bd}V^d)\mid_{t=0} 
\endalign
$$
and having the natural volume element
$$
\sqrt{\det~~ ^{(n)}g'} = (2^{1/2} V^t \sqrt{\det\mu})\mid_{t=0}. \tag3.23
$$ 

Since $X = \frac{\partial}{\partial x^{n}}$ in an agn chart adapted
to $X$, we have
$$
^{(n)}g'(X,X) = \,^{(n)}g'_{nn} = (2V^tV^t)\mid_{t=0} \tag3.24
$$
as a globally defined, nowhere vanishing function on $\Cal H$.  Using this
non-vanishing function as a conformal factor, we define a second
Riemannian metric $^{(n)}\tilde g$ on $\Cal H$, conformal to $^{(n)}g'$, by
setting
$$
^{(n)}\tilde g = (\frac{1}{V^{t}})^{2/n}\mid_{t=0}~~ ^{(n)}g'. \tag3.25
$$
The natural volume element of $^{(n)}\tilde g$ is thus given by
$$\align
\sqrt{\det~~ ^{(n)}\tilde g} &= (\frac{1}{V^{t}})\mid_{t=0} \sqrt{\det~~
^{(n)}g'} \tag3.26  \\
&= (2^{1/2} \sqrt{\det\mu})\mid_{t=0}
\endalign
$$
Computing the divergence of $X$ with respect to the metric $^{(n)}\tilde g$,
we find
$$\align
\nabla_{^{(n)}{\tilde g}}\cdot X &= \frac{1}{\sqrt{\det ~~^{(n)}\tilde g}}
\frac{\partial}{\partial x^{i}} (\sqrt{\det~~ ^{(n)}\tilde g} X^i) \tag3.27 \\ 
&= (\frac{1}{\sqrt{\det\mu}} \frac{\partial}{\partial x^{n}}
(\sqrt{\det\mu}))\mid_{t=0} \\
&= 0
\endalign
$$
which vanishes by virtue of the result discussed in the previous section (i.e., by virtue of the invariance of the
transversal metric $\overset{\circ}\to\mu_{ab}$ relative to the flow along $X$). Equation (3.27) can be equivalently expressed as
$$
\Cal L_X (\sqrt{\det~~ ^{(n)}\tilde g}) = 0 \tag3.28
$$
where $\Cal L$ signifies the Lie derivative. Thus the volume element
of $^{(n)}\tilde g$ is preserved by the flow along $X$.

It follows from the above that if $\{ f^\lambda\mid\lambda\in\bold
R\}$ is the one-parameter family of diffeomorphisms of $\Cal H$ generated
by $X$ and if $D$ is any measurable region of $\Cal H$ with volume
(relative to $^{(n)}\tilde g$) $\text{vol}(D)$ then $\text{vol}(f^\lambda D)
= \text{vol}(D)~~ \forall~~ \lambda\in\bold R$.  Since $\Cal H$ is compact and
$f^\lambda$ is volume preserving, the Poincar\'e recurrence theorem
may be applied and has the following consequences.  Let $p$ be a
point of $\Cal H$ and $U$ be any neighborhood of $p$ and, for any
$\lambda_0\ne 0$, consider the sequence of iterates
$f^{n\lambda_{0}}$ (for $n=1,2,\ldots$) of $f\equiv f^{\lambda_{0}}$
and the corresponding sequence of (equal volume) domains $U, fU,
f^2U, \ldots, f^nU, \ldots$.  Poincar\'e's theorem shows that there
always exists an integer $k > 0$ such that $f^kU$ intersects $U$ and
thus that, in any neighborhood $U$ of $p$, there always exists a
point $q$ which returns to $U$ under the sequence of mappings
$\{f^n\}$.

The above results together with those of the previous subsection show
that any point $p\in\Cal H$ eventually returns to an arbitrarily small
neighborhood of $p$ (after first leaving that neighborhood) when
followed along the flow of $X$.  The reason is that since, by
construction, $X = \frac{\partial}{\partial x^n}$ has no zeros on $\Cal H$, every point $p\in\Cal H$ flows
without stagnation along the integral curves of $X$, first leaving
sufficiently small neighborhoods of $p$ and then, by Poincar\'e
recurrence, returning arbitrarily closely to $p$.

It may happen that a point $p$ may actually flow back to itself, in
which case the generator it lies on is closed but, for the generic
points of interest here, the generators will not be closed and the
flow will only take $p$ back arbitrarily closely to itself.
\vskip .25in

\subhead 3.4.  A Connection on $\Cal H$ and some associated ``ribbon
arguments'' \endsubhead
\vskip .10in

Let $^{(n+1)}Y$ and $^{(n+1)}Z$ be any two smooth vector fields on
($^{(n+1)}U,g$) which are tangent to $\Cal H$ (i.e., for which
$^{(n+1)}Y^t\mid_{t=0}~ =~ ^{(n+1)}Z^t\mid_{t=0} = 0$ in an arbitrary
gaussian null  coordinate chart).  Then, computing the covariant
derivative $\nabla_{(n+1)_{Y}}~~ ^{(n+1)}Z$, determined by the spacetime
metric $g$, observe that the resulting vector field is automatically
also tangent to $\Cal H$ as a consequence of the invariance property of
the transversal metric which was derived in Section 3.2 (i.e., of the
result that $\overset{\circ}\to\mu_{ab,n} = 0$).  This fact, which
corresponds to the vanishing of the connection components
$\Gamma^t_{ij}\mid_{t=0}$ (for $i,j = 1,2,\ldots ,n$) of the metric $g$, in turn implies that
$\Cal H$ is totally geodesic (i.e., that every geodesic of $g$ initially
tangent to $\Cal H$ remains in $\Cal H$ through its entire interval of
existence).

If $Y$ and $Z$ designate the vector fields on $\Cal H$ induced by
$^{(n+1)}Y$ and $^{(n+1)}Z$ respectively, then we can, by virtue of the
above remarks, define a connection $^{(n)}\Gamma$ on $\Cal H$ by means of
the following defining formula for covariant differentiation
$$
^{(n)}\nabla_Y Z\equiv (\nabla_{(n+1)_{Y}} ~~^{(n+1)}Z)\mid_\Cal H.\tag3.29
$$
Here the right hand side symbolizes the vector field naturally
induced on $\Cal H$ by \newline $\nabla_{(n+1)_{Y}}~~ ^{(n+1)}Z$. A straightforward
computation in gaussian null coordinate charts (restricted to $\Cal H$)
shows that
$$
(^{(n)}\nabla_Y Z)^k = Y^j Z^k_{,j} + ~^{(n)}\Gamma^k_{ij} Z^i Y^j \tag3.30
$$
where
$$
^{(n)}\Gamma^k_{ij} = \Gamma^k_{ij}\mid_{t=0}\tag3.31
$$
and where $\Gamma^\alpha_{\beta\gamma}$ are the Christoffel symbols of
$g_{\alpha\beta}$.  The components of $^{(n)}\Gamma$ are given
explicitly by
$$\align
&^{(n)}\Gamma^n_{nn} = -\frac{1}{2} \overset{\circ}\to\varphi_{,t},\quad
^{(n)}\Gamma^n_{an} = -\frac{1}{2} \overset{\circ}\to\beta_{a,t} \tag3.32 \\
&^{(n)}\Gamma^n_{ab} = -\frac{1}{2} \overset{\circ}\to\mu_{ab,t}, \quad
^{(n)}\Gamma^d_{nn} = 0 \\
&^{(n)}\Gamma^d_{na} = 0, \quad
^{(n)}\Gamma^d_{ab} =~ ^{(n-1)}\overset{\circ}\to\Gamma^d_{ab}  \\
\endalign
$$
where $^{(n)}\Gamma^k_{ij} =~~ ^{(n)}\Gamma^k_{ji}$ and where the
$^{(n-1)}\overset{\circ}\to\Gamma^d_{ab}$ are the Christoffel symbols
of the invariant transversal metric $\overset{\circ}\to\mu_{ab}(x^c)$.

A similar calculation shows that if $^{(n+1)}\Omega$ is a one-form on
$(^{(n+1)}U,g)$ and $\Omega$ its pull-back to $\Cal H$ then the pull-back of
$\nabla_{(n+1)_{Y}}~ ^{(n+1)}\Omega$ is given by $^{(n)}\nabla_Y
~\Omega$ where, as expected,
$$
(^{(n)}\nabla_Y ~\Omega)_i = Y^j\Omega_{i,j} - ~~^{(n)}\Gamma^k_{ij}
Y^j\Omega_k. \tag3.33
$$

Now, recall the fixed vector field $X$ which was introduced in Section 3.1 and, for simplicity, work in agn charts adapted to $X$ so that $X = \frac{\partial}{\partial x^{n}}$. For an arbitrary vector field
$Z$ defined on $\Cal H$ we find, by a straightforward computation, that
$$
^{(n)}\nabla_Z ~X = (\omega_X(Z))X \tag3.34
$$
where $\omega_X$ is a one-form given, in the agn charts adapted to $X$, by
$$
\omega_X = - \frac{1}{2} \overset{\circ}\to\varphi_{,t} dx^n - \frac{1}{2}
\overset{\circ}\to\beta_{a,t} dx^a.\tag3.35
$$

The exterior derivative of $\omega_X$ is readily found to be
$$\align
d\omega_X = &-\frac{1}{2}(\overset{\circ}\to\varphi_{,ta} -
\overset{\circ}\to\beta_{a,tn}) dx^a\wedge dx^n \tag3.36 \\
&-\frac{1}{2} \overset{\circ}\to\beta_{a,tb} dx^b\wedge dx^a.
\endalign
$$
However, the Einstein equation $R_{nb} = 0$, restricted to $\Cal H$ and
reduced through the use of $\overset{\circ}\to\mu_{ab,n} = 0$,
becomes (c.f. Equation (3.2) of Reference \cite{9}):
$$
\overset{\circ}\to\varphi_{,ta} - \overset{\circ}\to\beta_{a,tn} = 0. \tag3.37
$$
Thus $d\omega_X$ reduces to
$$
d\omega_X = -\frac{1}{2} \overset{\circ}\to\beta_{a,tb} dx^b\wedge dx^a. \tag3.38
$$

In subsequent sections, we shall be studying integrals of the form
$$
\int_\gamma\omega_X = \int_\gamma (-\frac{1}{2}
\overset{\circ}\to\varphi_{,t})dx^n \tag3.39
$$
along segments $\gamma$ of integral curves of $X$.  We shall be
interested in comparing the values of these integrals for nearby
integral curves.  For that purpose, the following sort of \it ribbon
argument \rm will prove indispensable.

Let $p$ and $p'$ be any two points of $\Cal H$ which can be connected by a
smooth curve which is everywhere transversal to the flow of $X$.  Let
$c: I\to N$ be such a curve defined on the interval $I = [a,b]$ with
$c(a) = p$ and $c(b) = p'$ and let $\ell :I\to\bold R$ be a smooth,
strictly positive function on $I$.  Now consider the strip or
\it ribbon \rm generated by letting each point $c(s)$ of the curve $c$
flow along $X$ through a parameter distance $\ell (s)$ (i.e., through
a lapse of $\ell (s)$ of the natural curve parameter defined by $X$).
 This construction gives an immersion of the ribbon
$$
r = \{ (s,t)\in\bold R^2\mid s\in I, 0\leq t\leq\ell (s)\}
$$
into $\Cal H$ which consists of connected segments of integral curves of
$X$.  In  particular, the integral curves starting at $p$ and $p'$
form the \it edges \rm of the ribbon whereas the initial curve $c$
together with its image after flow along $X$ form the \it ends \rm of
the ribbon. 

If $i:r\to\Cal H$ is the mapping which immerses $r$ in $\Cal H$ according to
the above construction and $i^* \omega_X$ and $i^* d\omega_X$ are the
pull-backs of $\omega_X$ and $d\omega_X$ to $r$ respectively, then one sees
from Equation (3.38) and the tangency of the ribbon to the integral curves
of $X$, that $i^* d\omega_X = 0$.

Therefore, by means of Stokes theorem, we get
$$
\int_{\partial r}\omega_X = \int_r d\omega_X = 0 \tag3.40
$$
for any ribbon of the type described above.  Thus if $\gamma$ and
$\gamma'$ designate the two edges of $r$ (starting from $s = a$ and
$s = b$ respectively and oriented in the direction of increasing $t$)
and if $\sigma$ and $\sigma'$ designate the two \it ends \rm of $r$
defined by $\sigma = \{ (s,0)\mid s\in I\}$ and $\sigma' = \{ (s,\ell
(s))\mid s\in I\}$  respectively (and oriented in the direction of
increasing $s$) then we get, from $\int_{\partial r} \omega_X = 0$, that
$$
\int_\gamma \omega_X - \int_{\gamma'}\omega_X = \int_\sigma \omega_X -
\int_{\sigma'}\omega_X . \tag3.41
$$

Equation (3.41) will give us a means of comparing $\int_\gamma\omega_X$
with $\int_{\gamma'}\omega_X$ provided we can estimate the contributions
to $\int_{\partial r}\omega_X$ coming from the ends of the ribbon.  As a
simple example,  suppose (as we did in Reference \cite{9}) that every integral
curve of $X$ is closed and choose $r$ and $i:r\to\Cal H$ so that the
image of $r$ in $\Cal H$ consists of a ribbon of simply closed curves.  In
this case, the end contributions cancel and we get that $\int_\gamma
\omega_X = \int_{\gamma'}\omega_X$.  This result played a key role in the
arguments of Reference \cite{9}.

\newpage

\head IV. Nondegeneracy and geodesic incompleteness\endhead
\vskip .10in

In this section we shall show, using a ribbon argument, that each null geodesic generator of $\Cal H$ is either complete in both directions (the ``degenerate'' case) or else that each generator is incomplete in one direction (the non-degenerate case).  More precisely, we shall prove that if any single generator $\gamma$ is incomplete in a particular direction (say that defined by $X$) then every other generator of the (connected) hypersurface $\Cal H$ is necessarily incomplete in the same direction.  It will then follow that if any generator  is complete in a particular direction, then all must be since otherwise one could derive a contradiction from the first result.  We shall see later that, in the non-degenerate case, the generators which are all incomplete in one direction (say that of $X$) are however all complete in the opposite direction (that of $-X$).  For black hole spacetimes (whose compactified horizons provide the hypersurfaces $\Cal H$) the degenerate case will turn out to correspond to that of vanishing surface gravity (i.e., to that of extremal black holes) whereas the non-degenerate case will prove to correspond to non-vanishing, but constant surface gravity defined on the horizons.

As usual we work in adapted charts for $\Cal H$.  For the calculations to follow however, it is convenient to work with charts induced from adapted charts on the covering space $\Cal H\approx\Sigma\times\bold R$ for which the $\{ x^a\mid a = 1,\ldots, n-1\}$ are constant along any given generator and the range of the ``angle'' coordinate $x^n$ is unwrapped to cover the interval $(-\infty, \infty)$.  Projected back to $\Cal H$ these induce families of charts $\{ x^n, x^a \}, \{ x^{n'}, x^{a'}\}$, etc. related, on their regions of overlap, by analytic transformations of the form
$$\align
&x^{n'} = x^n + \text{constant} \tag4.1 \\
&x^{a'} = f^a(x^1, \ldots, x^{n-1}).
\endalign
$$
By working on the covering space we simplify the notation by keeping the $\{ x^a\}$ constant and letting $x^n$ range continuously over $(-\infty, \infty)$ in following a given generator as it repeatedly sweeps through the leaves of the chosen foliation of $\Cal H$.  However, one should keep in mind that this is just an artifice to represent calculations carried out on the {\it compact} manifold $\Cal H$ in a simplified notation since the compactness of $\Cal H$ will play a key role in the arguments to follow.

Consider a null generator of $\Cal H$ developed from ``initial'' conditions specified at a point $p\in \Cal H$ having coordinates $\{ x^n (p) = \overset o\to x^n, x^a(p) = \overset{o}\to x^a\}$. The affine parametrization of this generator is determined by solving the geodesic equations which, for the class of curves in question, effectively reduce to
$$\align
&\frac{d^2x^n}{d\eta^{2}} - \frac{\overset{o}\to\varphi_{,t}}{2}  (x^n, x^a) \left(\frac{dx^n}{d\eta}\right)^2 = 0 \tag4.2 \\
&x^a(\lambda) = \overset{o}\to x^a = \,\,\text{constant}
\endalign
$$
where $\eta$ is an affine parameter. To complete the specification of initial conditions one needs, of course, to give an initial velocity $\frac{dx^n}{d\eta}\mid_{\overset{o}\to\eta}$ (taking $\frac{dx^a}{d\eta}\mid_{\overset{o}\to\eta} = 0$).

Solving the first order equation  
$$
\frac{dv}{d\eta} = \frac{\overset{o}\to\varphi_{,t}}{2} \,\, v^2 \tag4.3
$$
for $v := \frac{dx^n}{d\eta}$ to get an integral formula for $v$ and then integrating $\frac{d\eta}{dx^n} = \frac{1}{v}$ with respect to $x^n$ one derives an expression for the affine length of a segment of this null geodesic defined on the interval $[\overset{o}\to x^n, x^n]$: 
$$\align
&\eta (x^n,\overset{o}\to x^a) - \overset{o}\to\eta (\overset{o}\to x^n, \overset{o}\to x^a) \tag4.4 \\
&=\frac{1}{(\frac{dx^n}{d\eta})}\bigm|_{\overset{o}\to\eta (\overset{o}\to x^{n},\overset{o}\to x^{a})} \int\limits^{x^{n}}_{\overset{o}\to x^{n}} d\rho \quad \text{exp} [-\int\limits^\rho_{\overset{o}\to x^{n}} d\xi (\frac{\overset{o}\to\varphi_{,t}}{2} (\xi,\overset{o}\to x^a)] .
\endalign
$$
Thus incompleteness of this generator, in the direction of $X = \frac{\partial}{\partial x^{n}}$, would correspond to the existence of the limit
$$\align
&\lim\limits_{x^{n}\to\infty} \int\limits^{x^{n}}_{\overset{o}\to x^{n}} d\rho \quad\text{exp} [-\int\limits^\rho_{\overset{o}\to x^{n}} d\xi (\frac{\overset{o}\to\varphi_{,t}}{2} (\xi , \overset{o}\to x^a))] \tag4.5  \\
&\quad =(\frac{dx^{n}}{d\eta})\bigm|_{\overset{o}\to\eta (\overset{o}\to x^{n},\overset{o}\to x^a)} (\eta (\infty, \overset{o}\to x^a) - \overset{o}\to\eta (\overset{o}\to x^n, \overset{o}\to x^a)) < \infty
\endalign
$$
whereas completeness (in this direction)  would correspond to the divergence of this limit. Recalling Equation (3.39), note that the integral of the one-form $\omega_X$ along the segment $\gamma$ defined above is given by
$$
\int\limits_\gamma \omega_X = \int\limits^{x^{n}} _{\overset{o}\to x^{n}} (-\frac{1}{2} \overset{o}\to\varphi,_{t} (\xi , \overset{o}\to x^a))d\xi \tag4.6
$$
which thus provides an invariant representation of the basic integral arising in the above formulas.

Suppose that the generator ``beginning'' at $p\in \Cal H$ is incomplete in the direction of $X$. We want to establish convergence of the corresponding integral for any other generator of $\Cal H$.  Since incompleteness is an asymptotic issue (the relevant integrals being automatically finite on any compact domain of integration) there is no essential loss of generality in comparing only those generators that ``start'' in the slice defined by $p$.  Thus we want to consider generators ``beginning'' at points $q$, having $x^n (q) = \overset{o}\to x^n$, and establish their incompleteness by using a suitable ribbon argument.  Furthermore, to have a ``canonical'' way of defining our comparison ribbons it will be convenient to localize the calculations somewhat by first looking only at generators sufficiently near to the ``reference'' generator. Thus, given a point $p$ in the initial slice defined by $x^n(p) = \overset{o}\to x^n$, we consider only those points $q$ lying in this slice which, additionally, lie within a closed geodesic ball (relative to the invariant transversal metric $\mu$ induced on this slice) centered at $p$ and contained within a normal neighborhood of this point.  Any such $q$ can be connected to $p$ by a unique geodesic lying within this geodesic ball and such points can be conveniently labeled by normal coordinates defined at $p$ (i.e., the points of a corresponding, closed ball in the tangent space to the slice at $p$).

The unique geodesic connecting $q$ to $p$ provides a canonical ``starting end'' to our comparison ribbon for geodesics emanating from points $p$ and $q$ (in the direction of $X$) and, from invariance of the transversal metric along the flow of $X$, we get an isometric image of this connecting geodesic induced on any subsequent slice traversed along the flow.

Let $\gamma$ be the segment of the null generator beginning at $p$ and defined on the interval $[\overset{o}\to x^n, x^n]$, for some $x^n > \overset{o}\to x^n$, and let $\gamma'$ be a corresponding segment of the generator beginning at $q$ and defined on the same interval. From the argument given in Section 3.4 it follows that
$$
\int\limits_\gamma \omega_X - \int\limits_{\gamma'} \omega_X = \int\limits_\sigma \omega_X  - \int\limits_{\sigma'} \omega_X \tag4.7
$$
where $\sigma$ is the geodesic end defined in the starting slice and $\sigma'$ its isometric image at the ending slice.

For fixed $p$ the integral $\int_\sigma \omega_X$ varies continuously with $q$ as $q$ ranges over a compact set (the closed geodesic ball centered at $p$ described above) and thus is bounded for all $q$ in this ball. Furthermore the integral $\int_{\sigma'} \omega_X$ varies continuously with $q$ and $x^n$ but, as $x^n$ increases, the image of $p$ under the flow ranges only over (some subset of) the {\it compact} manifold $\Cal H$ whereas the image of $q$ remains always a fixed geodesic distance from the image of $p$ in the corresponding slice. Since the product of $\Cal H$ with this (closed) ball is compact the continuously varying integral $\int_{\sigma'} \omega_X$ (regarded as a function of $q$ and $x^n$ for fixed $p$) is necessarily bounded no matter how large the ``unwrapped'' coordinate $x^n$ is allowed to become.  

It follows from the forgoing that for any fixed $p$ and $q$ as above, there exists a {\it bounded}, continuous (in fact analytic) real-valued function $\delta_{p,q}(x^n)$ such that
$$
\int\limits_{\gamma'} \omega_X = \int\limits_\gamma \omega_X + \delta_{p,q}(x^n)\tag4.8
$$
for arbitrary $x^n > \overset{o}\to x^n$.  But this implies that
$$\align
&\int\limits_{\overset{o}\to x^{n}}^{x^{n}} d\rho \,\, \text{exp} [-\int\limits_{\overset{o}\to x^{n}}^\rho \frac{\overset{o}\to\varphi_{,t}}{2} (\xi, \overset{o}\to x^a(q))d\xi] \\
&= \int\limits_{\overset{o}\to x^{n}}^{x^{n}} d\rho \,\, \text{exp} [-\int\limits_{\overset{o}\to x^{n}}^\rho \frac{\overset{o}\to\varphi_{,t}}{2} (\xi, \overset{o}\to x^a(p))d\xi + \delta_{p,q} (\rho)] \tag4.9 \\
&= \int\limits_{\overset{o}\to x^{n}}^{x^{n}} d\rho \,\, \text{exp}[\delta_{p,q}(\rho)] \,\,\text{exp} [-\int\limits_{\overset{o}\to x^{n}}^\rho \frac{\overset{o}\to\varphi_{,t}}{2} (\xi, \overset{o}\to x^a(p))d\xi]  .
\endalign
$$
From the boundedness of $\delta_{p,q}$
$$
- \infty < b_1 \leq \delta_{p,q}(\rho)\leq b_2 < \infty,
\forall\rho\in [\overset{o}\to x^n, \infty) \tag4.10
$$
it follows that
$$\align
&e^{b_{1}} \int\limits_{\overset{o}\to x^{n}}^{x^{n}} d\rho \,\, \text{exp}[-\int\limits_{\overset{o}\to x^{n}}^\rho \frac{\overset{o}\to\varphi_{ ,t}}{2} (\xi, \overset{o}\to x^a(p))d\xi] \tag4.11 \\
&\leq \int\limits_{\overset{o}\to x^{n}}^{x_{n}} d\rho \,\,\text{exp}[-\int\limits_{\overset{o}\to x^{n}}^\rho \frac{\overset{o}\to\varphi_{ ,t}}{2} (\xi, \overset{o}\to x^a(q))d\xi] \\
&\leq e^{b_{2}} \int\limits_{\overset{o}\to x^{n}}^{x^{n}} d\rho \,\, \text{exp}[-\int\limits_{\overset{o}\to x^{n}}^\rho \frac{\overset{o}\to\varphi _{,t}}{2} (\xi, \overset{o}\to x^a(p))d\xi]
\endalign
$$ 
$\forall x^n\in [\overset{o}\to x^n ,\infty)$. But this implies that if the limit
$$
\lim\limits_{x^{n}\to\infty} \int^{x^{n}}_{\overset{o}\to x^{n}} d\rho \,\, \text{exp}[-\int^\rho_{\overset{o}\to\x^{n}} \frac{\overset{o}\to\varphi _{,t}}{2} (\xi, \overset{o}\to x^a(p))d\xi]
$$
exists, then so must the limit of the monotonically increasing function \newline $\int^{x^{n}}_{\overset{o}\to x^{n}} d\rho \,\, \text{exp}[-\int^\rho_{\overset{o}\to\x^{n}} \frac{\overset{o}\to\varphi_{ ,t}}{2} (\xi, \overset{o}\to x^a(q))d\xi]$ exist as $x^n\to\infty$. Conversely, if the affine length of $\gamma$ diverges, then so must that of $\gamma'$ by virtue of the forgoing bounds.

So far we have only considered geodesics starting within a geodesic ball centered at a point $p$ in the initial slice.  But one can always cover this (compact, connected) slice with a finite number of such balls and thus deduce that either all the geodesic generators of $\Cal H$ are incomplete in the direction of $X$, or else they are all complete in this direction. Clearly the same argument can be applied in the opposite direction (i.e., that of $-X$) with a corresponding conclusion. However, as we shall see later, the non-degenerate case will always be characterized by generators that are all incomplete in one direction but complete in the opposite direction, whereas the degenerate case will be characterized by generators that are complete in both directions.

\newpage

\head V. A candidate vector field in the non-degenerate case\endhead
\vskip .10in

In this section, we focus on the non-degenerate case and, if necessary, change the sign of $X$ so that it points in a direction of incompleteness for the null generators of $\Cal H$. We now define a vector field $K$ on $\Cal H$, also tangent to the generators of this hypersurface, by setting $K = uX$ where $u$ is a positive real-valued function on $\Cal H$ chosen so that, for any point $p\in\Cal H$, the null generator determined by the initial conditions $(p, K(p) = u(p) X(p))$ has a fixed (i.e., independent of $p$) future affine length given by $\frac{2}{k}$ where $k$ is a constant $> 0$.  At the moment there is no preferred normalization for $k$ so we choose its value arbitrarily. We shall see later however, that a (non-degenerate) stationary black hole has a non-vanishing constant surface gravity defined globally on its horizon and that the value of this quantity naturally determines a scale for $k$.

From Equation (4.5) upon putting $(\eta (\infty, \overset{o}\to x^a) - \overset{o}\to\eta (\overset{o}\to x^n, \overset{o}\to x^a)) = \frac{2}{k}$, we see that $u(x^n, x^a)$ is necessarily given by
$$
u(x^n,x^a) = \frac{k}{2} \int\limits^\infty_{x^{n}} d\rho ~ \text{exp}[-\int\limits^{\rho}_{x^{n}} \frac{\overset{o}\to\varphi,_{t}}{2} (\xi, x^a)d\xi]. \tag5.1
$$
By the results of the previous section, the needed integral converges for every generator and clearly $u > 0$ on $\Cal H$.  What is not clear however, in view of the limiting procedure needed to define the outer integral over a semi-infinite domain, is whether $u$ is in fact analytic and we shall need to prove that it is.  We shall do this below by showing that a sequence $\{ u_i : \Cal H\to\bold R^+\mid i = 1,2,\ldots\}$ of analytic ``approximations'' to $u$ defined by
$$
u_i(x^n,x^a) = \frac{k}{2} \int\limits^{x^{n}+is^{*}}_{x^{n}} d\rho ~ \text{exp}[-\int\limits^\rho_{x^{n}} \frac{\overset{o}\to\varphi,_{t}}{2} (\xi, x^a)d\xi] \tag5.2
$$
where $s^*$ is the recurrence time introduced in Section 3.1 does indeed have an analytic limit as $i\to\infty$.  

For the moment however, let us assume that we know that $K$ is analytic and introduce new agn coordinates $\{ x^{n'}, x^{a'},t'\}$ which are adapted to $K$ rather than to $X = \frac{\partial}{\partial x^{n}}$. Thus we seek a transformation of the form $\{x^{n'} = h(x^n, x^a), x^{a'} = x^a\}$ which yields $K = \frac{\partial}{\partial x^{n'}}$. A straightforward calculation shows that $h$ must satisfy
$$
\frac{\partial h(x^{n},x^{a})}{\partial x^{n}} = \frac{1}{u(x^{n},x^{a})} = \{\frac{1}{\frac{k}{2} \int^\infty_{x^n} d\rho ~ \exp[-\int^\rho_{x^{n}} d\xi \frac{\overset{o}\to\varphi_{,t}}{2}(\xi, x^{a})]}\} \tag5.3
$$
which, since the denominator is analytic by assumption and non-vanishing, yields an analytic $h$ upon integration.

As was shown in Sect. IIIA of Reference [9], a transformation of the above type connects the primed and unprimed metric functions $\overset{o}\to\varphi_{,t}$ and $\overset{o}\to\varphi'_{,t'}$ via
$$
2 \frac{\partial}{\partial x^n}\left(\frac{\partial h}{\partial x^n}\right) + \frac{\partial h}{\partial x^n} \overset{o}\to\varphi_{,t} = \left(\frac{\partial h}{\partial x^n}\right)^2 \, \overset{o}\to\varphi'_{,t'} . \tag5.4
$$
Computing $\frac{\partial^2h}{\partial x^{n2}}$ from Equation (5.3) above and substituting this and 
$\frac{\partial h}{\partial x^n}$ into the above formula one finds that the transformed metric has
$$
\overset{o}\to\varphi'_{,t'} = k = \text{constant} \tag5.5
$$
throughout any agn chart adapted to $K$.  This argument is somewhat the reverse of that given in Reference [9], for the case of closed generators, wherein we set $\overset{o}\to\varphi'_{, t'} = k$ and solved Equation (5.4) for $\frac{\partial h}{\partial x^n}$ and then $h$.

In the new charts one still has $\overset{o}\to\varphi' = \overset{o}\to\beta'_\alpha = 0$ since these hold in any agn coordinate system and, upon repeating the argument of Section 3.2 above, with $K$ in place of $X$, we obtain 
$\overset{o}\to\mu'_{a'b',n'} = 0$ as well.  Now evaluating the Einstein equation $R_{nb} = 0$ at $t = t' = 0$ and using the foregoing, together with the new result that $\overset{o}\to\varphi'_{,t'} = k$ in the primed charts, one finds that $\overset{o}\to\beta'_{b',t',n'} = 0$.

Deleting primes to simplify the notation, we thus find that in agn charts adapted to $K$, the metric functions obey
$$
\overset{o}\to\varphi = \overset{o}\to\beta_a = \overset{o}\to\mu_{ab,n} = 0,\,\, \overset{o}\to\varphi_{,t} = \text{constant} \neq 0, (\overset{o}\to\beta_{a,t})_{,n} = 0. \tag5.6
$$
These are the main results we shall need for the inductive argument of Section VII to prove that there is a spacetime Killing field $Y$ such that $Y\mid_\Cal H = K$.

Referring to Equation (4.4) and evaluating the integrals in the new charts in which $\overset{o}\to\varphi_{,t} = k = \text{constant} > 0$ one sees easily that though the null generators are all incomplete towards the ``future'' they are in fact all complete towards the ``past'' (where here future and past designate simply the directions of $K$ and $-K$ respectively).  It may seem strange at first glance to say that any generator could have a fixed future affine length $(=\frac{2}{k})$ no matter where one starts along it, but the point is that this length is here always being computed from the geodesic initial conditions $(p, K(p))$.  If one starts with say $(q, K(q))$ and later reaches a point $p$ on the same generator, then the tangent to the (affinely parametrized) geodesic emanating from $q$ will not agree with $K(p)$ but will instead equal $c K(p)$ for some constant $c > 1$,  Only upon ``restarting'' the generator with the initial conditions $(p, K(p))$ will it be found to have the same future affine length that it had when started instead from $(q, K(q))$.  Indeed, if the tangent to an affinely parametrized geodesic did not increase relative to $K$ then the generator could never be incomplete on a compact manifold $\Cal H$ where the integral curves of a vector field $K$ are always complete.

Let us now return to the question of the analyticity of the ``scale factor'' $u(x^n,x^a)$.  First note that, upon combining Equations (5.3), (5.4) and (5.5), $u$ satisfies the linear equation with analytic coefficients
$$
\frac{\partial u}{\partial x^n} - \frac{\overset{o}\to\varphi_{,t}}{2} \,\,  u = - \frac{k}{2} \tag5.7
$$
provided one takes, as initial condition specified at some $\overset{o}\to x^n$,
$$
u(\overset{o}\to x^n, x^a) = \frac{k}{2} \int\limits^\infty_{\overset{o}\to x^n} d\rho \,\, \text{exp}[-\int\limits^\rho_{\overset{o}\to x^n} \frac{\overset{o}\to\varphi_{,t}}{2} (\xi, x^a)d\xi ]. \tag5.8
$$
More precisely, using an appropriate integrating factor for Equation (5.7), namely \newline $\text{exp}[-\int^{x^{n}}_{\overset{o}\to x^n} d\xi \frac{\overset{o}\to\varphi_{,t}}{2} (\xi, x^a)]$, one easily shows that the solution to Equation (5.7) determined by the initial condition (5.8) is given by Equation (5.3).  But Equation (5.7) can be viewed as a (linear, analytic) partial differential equation to which the Cauchy Kowalewski theorem applies \cite{19} and guarantees the analyticity of the solution on domains corresponding (because of linearity) to those of the coefficients (in this case $\overset{o}\to\varphi_{,t} (x^n,x^a))$ {\it provided} that the initial condition $u(\overset{o}\to x^n, x^a)$ is analytic with respect to the $\{ x^a\}$.  In other words, our problem reduces to that of proving that Equation (5.8) for fixed $\overset{o}\to x^n$, defines an analytic function of the $\{x^a\}$.  Thus we only need to show that the sequence of ``approximations'' 
$$\align
&u_i(\overset{o}\to x^n, x^a) := \frac{k}{2} \int\limits^{\overset{o}\to x^{n} + is^{*}}_{\overset{o}\to x^{n}} d\rho \,\, \text{exp} [-\int\limits^\rho_{\overset{o}\to x^{n}} \frac{\overset{o}\to\varphi_{,t}}{2} (\xi, x^a)d\xi],\tag5.9 \\
&i = 1, 2, \ldots
\endalign
$$
converges to an analytic function of the $\{x^a\}$ for fixed $\overset{o}\to x^n$.

However, a (pointwise) convergent sequence of analytic functions could easily converge to a limit which is not even continuous much less analytic.  On the other hand, the set of continuous functions on a compact manifold forms a Banach space with respect to the $C^0$ norm (uniform convergence) so that one could hope at least to establish the continuity of the limit by showing that the sequence $\{u_i (\overset{o}\to x^n, x^a)\}$ is Cauchy with respect to this norm.

A much stronger conclusion is possible however, if one first complexifies the slices $x^n = \text{constant}$ (which are each diffeomorphic to the compact manifold $\Sigma$ defined previously) and extends the analytic metric functions defined on $\Cal H$ to holomorphic functions defined on this complex ``thickening'' of $\Cal H$ in the $\{ x^a\}$ directions.  The space of holomorphic functions on such a complex manifold (with boundary) forms a Banach space with respect to the $C^0$ norm so that the limit of any Cauchy sequence of holomorphic functions (which extend continuously to the boundary) will in fact be holomorphic and not merely continuous \cite{20, 21}.  In the following section, we shall define a certain complex `thickening' of $\Cal H$ with respect to all of its dimensions (a so-called `Grauert tube') but then, in view of the discussion in the preceding paragraph, restrict the integration variable $x^n$ to real values so that, in effect, only the leaves of the foliation of  $\Sigma\times \bold S^1$ are thickened.

Let us temporarily remain within the real analytic setting to sketch out the basic idea of the argument to be given later in the holomorphic setting.  This detour, though it cannot yield more than the continuity of $u(\overset{o}\to x^n, x^a)$ in the $\{ x^a\}$ variables, will be easier to understand at a first pass and will require only straightforward modification for its adaptation to the holomorphic setting.

For any point $p$ in the slice determined by $x^n(p) = \overset{o}\to x^n$ the monotonically increasing, convergent sequence of real numbers
$$\align
u_i(\overset{o}\to x^n, x^a(p)) &= \frac{k}{2} \int\limits^{\overset{o}\to x^{n} + is^{*}}_{\overset{o}\to x^{n}} d\rho  \,\, \text{exp} [-\int\limits^\rho_{\overset{o}\to x^{n}} d\xi \,\, \frac{\overset{o}\to\varphi_{,t}}{2} (\xi, x^a(p))] \tag5.10 \\
&i = 1,2, \ldots 
\endalign
$$
is clearly a Cauchy sequence which converges to $u(\overset{o}\to x^n, x^a(p))$.  Thus for any $\varepsilon' > 0$ there exists a positive integer $N$ such that
$$
\mid u_m(\overset{o}\to x^n, x^a(p)) - u_\ell (\overset{o}\to x^n, x^a(p))\mid < \varepsilon' \quad  \forall \,\, m, \ell > N. \tag5.11
$$
Now consider an arbitrary point $q$ in the initial slice (i.e., having $x^n(q) = \overset{o}\to x^n$) that lies within a closed geodesic ball in this slice which is centered at $p$ (i.e., a ball of the type used in the ribbon argument of the previous section).  By the ribbon arguments given in this last section, one easily finds that
$$\align
&\mid u_m(\overset{o}\to x^n, x^a(q)) - u_\ell (\overset{o}\to x^n, x^a(q))\mid \\
&= \biggm|\frac{k}{2}\int\limits^{\overset{o}\to x^{n} + ms^{*}}_{\overset{o}\to x^{n}+\ell s^{*}} \,\,d\rho \,\, \text{exp}[-\int\limits^\rho_{\overset{o}\to x^n} \,\, d\xi \, \frac{\overset{o}\to\varphi_{,t}}{2} (\xi, x^a(q))]\biggm| \\
&= \biggm|\frac{k}{2} \int\limits^{\overset{o}\to x^{n} + ms^{*}}_{\overset{o}\to x^{n}+\ell s^{*}} \,\,d\rho \,\, \text{exp}[\delta_{p,q}(\rho)] \text{exp}[-\int\limits^\rho_{\overset{o}\to x^n} \,\, d\xi \, \frac{\overset{o}\to\varphi_{,t}}{2} (\xi, x^a(p))]\biggm|  \tag5.12 \\
&\leq e^{b_{2}} \biggm|\frac{k}{2} \int\limits^{\overset{o}\to x^{n} + ms^{*}}_{\overset{o}\to x^{n}+\ell s^{*}} \,\,d\rho \,\, \text{exp}[-\int\limits^\rho_{\overset{o}\to x^n} \,\, \frac{\overset{o}\to\varphi_{,t}}{2} (\xi, x^a(p))d\xi]\biggm| \\
&= e^{b_{2}} \bigm| u_m(\overset{o}\to x^n, x^a(p)) - u_\ell (\overset{o}\to x^n, x^a(p))\bigm|
\endalign
$$ 
for all $q$ in this ball where $b_2$ is a constant that depends upon $p$ and the radius of the chosen ball.  Thus for any $\varepsilon > 0$ we get by choosing $\varepsilon'  = e^{-b_{2}}\varepsilon$ in Equation (5.11), that
$$
\bigm| u_m (\overset{o}\to x^n, x^a(q)) - u_\ell (\overset{o}\to x^n, x^a(q))\bigm| \,\, < \varepsilon \quad \forall \, m, \ell > N \tag5.13
$$
and for all $q$ in the compact set defined by the chosen (closed) geodesic ball.  Thus the sequence of (real-valued) continuous functions $\{ u_m(\overset{o}\to x^n, x^a(q))\mid m = 1,2,\ldots\}$ defined on this ball is a Cauchy sequence relative to the $C^0$-norm and hence its limits $u(\overset{o}\to x^n, x^a(q))$ is necessarily continuous.  By covering the initial slice by a finite collection of such balls, we deduce that $u(\overset{o}\to x^n, x^a(q))$ is globally continuous on the initial slice.

\newpage

\head VI. Analyticity of the candidate vector field \endhead
\vskip .10in

Let us define a Riemannian metric $g$ on the horizon manifold $\Cal H$ by writing on $\tilde\Cal H\approx\Sigma\times\bold R$,
$$\align
g &= g_{ij} (x^1,\ldots. x^{n-1}) dx^i\otimes dx^j\tag6.1 \\
&= dx^n\otimes dx^n + \mu_{ab} (x^1,\ldots, x^{n-1}) dx^a\otimes dx^b
\endalign
$$
and then, as before, identifying the slice at $x^n$ with that at $x^n + s^*$ via the aforementioned analytic isometry of $(\Sigma, \mu)$.  This metric is adapted to the chosen slicing of $\Cal H$ in that each $x^n = \text{constant}$ slice is a totally geodesic submanifold of $(\Cal H,g)$ and furthermore the integral curves of $X = \frac{\partial}{\partial x^{n}}$, which is evidently a Killing field of $g$, coincide with the geodesics of $(\Cal H,g)$ normal to the $x^n = \text{constant}$ slices.

There is a canonical way of complexifying a compact, analytic Riemannian manifold such as $(\Cal H,g)$ through the introduction of its so-called Grauert tubes \cite{22}.  One identifies $\Cal H$ with the zero section of its tangent bundle $T\Cal H$ and defines a map $\ell :T\Cal H\to\bold R$ such that $\ell (v)$ is the length of the tangent vector $v\in T\Cal H$ relative to the Riemannian metric $g$.  Then, for sufficiently small $s > 0$, the manifold (`Grauert tube' of thickness $s$)
$$
T^s\Cal H = \{ v\in T\Cal H \mid \ell (v) < s\} \tag6.2
$$
can be shown to carry a complex structure for which holomorphic coordinates $\{ z^i\}$ can be defined in terms of analytic coordinates $\{ x^i\}$ for $\Cal H$ by setting $z^k = x^k + iy^k$ where $y = y^k \frac{\partial}{\partial x^{k}}$ represents a vector in $T\Cal H$.  Analytic  transformations between overlapping charts for $\Cal H$ extend to holomorphic transformations between  corresponding charts for $T^s\Cal H$ provided that, as we have assumed, $\Cal H$ is compact and $s$ is sufficiently small.  For non-compact $\Cal H$ such a holomorphic thickening need not exist for any $s$, no matter how small, and further restrictions upon $(\Cal H,g)$ are in general needed in order to define its Grauert tubes. When defined, Grauert tubes have an anti-holomorphic involution $\sigma: T^s\Cal H\to T^s\Cal H$ given by $v\mapsto -v$.

From the special properties of the metric $g$ and its geodesics, it is easy to see that if $\{ x^a\mid a= 1,\ldots ,n-1\}$ are normal coordinates for $(\Sigma, \mu)$ centered at a point $q\in\Sigma$ (with, therefore, $x^a(q) = 0$) then, holding these constant along the flow of $X$ and, complementing them with the function $x^n$, we get normal coordinates \newline $\{ x^i\} = \{ (x^a, x^n)\mid a = 1,\ldots , n-1\}$ defined on a tubular domain in $\Cal H$ centered on the orbit of $X$ through $q$.  By shifting $x^n$ by an additive constant, one can of course arrange that the origin of these normal coordinates for this tubular domain lies at any chosen point along the orbit through $q$.  It follows from the aforementioned property of Grauert tubes that the functions
$$
\{ z^k\} = \{ (z^k = x^k + iy^k) \mid (y^n)^2 + \mu_{ab} (x^1,\ldots ,x^{n-1}) y^a y^b < s\} \tag6.3
$$
will provide holomorphic coordinates on a corresponding domain for $T^s\Cal H$.

In the application to follow, as already mentioned in the previous section, we shall set $y^n = 0$ and thus focus our attention on `thickenings' of $\Cal H$ of the restricted form $T^s\Sigma\times\bold S^1$ which are foliated by curves of the type
$$\align
z^a(\lambda) &= x^a(\lambda) + iy^a(\lambda) = \overset{o}\to x^a + i\overset{o}\to y^a \tag6.4 \\
&= \text{constant}, \\
z^n(\lambda) &= \overset{o}\to x^n + \lambda, \quad y^n(\lambda) = 0, 
\endalign
$$
with
$$
\mu_{ab} (\overset{o}\to x^1, \ldots, \overset{o}\to x^{n-1}) \overset{o}\to y^a\overset{o}\to y^b < s. \tag6.5
$$
The closure $\overline{T^s\Sigma\times\bold S^1}\approx \overline{T^s\Sigma}\times\bold S^1$, of this manifold results from attaching a boundary characterized locally by $\mu_{ab} (x^1,\ldots, x^{n-1})y^ay^b = s$ to $T^s\Sigma\times\bold S^1$ and will also play a role in the considerations to follow.

Analytic tensor fields defined on $\Cal H$ can always, in view of its compactness, be lifted to define holomorphic fields on thickenings of the type $T^s\Cal H$ which, furthermore, extend continuously to the boundary of $\overline{T^s\Cal H}$ provided $s > 0$ is taken to be sufficiently small.  The needed limitation on the size of $s$ arises from considering the radii of convergence of the local series representations of these fields on the original analytic manifold $\Cal H$ but, since it is compact, a finite collection of such representations suffices to define the field globally on $\Cal H$ and hence a choice of $s > 0$ is always possible so that a given field on $\Cal H$ extends holomorphically to $T^s\Cal H$.  Upon restricting such a field to the manifold $T^s\Sigma\times\bold S^1$, as defined by setting $y^n = 0$, one obtains a corresponding field that is holomorphic with respect to the $\{ z^a \mid a = 1, \ldots , n-1\}$, real analytic with respect to $x^n$ and which extends continuously to the boundary of $\overline{T^s\Sigma\times\bold S^1} \approx \overline{T^s\Sigma}\times\bold S^1$. From our point of view, the important thing is that such fields form a Banach space with respect to the $C^0$ norm and hence a Cauchy sequence with respect to this norm will necessarily converge to a holomorphic field with respect to the $\{ z^a\}$.

To carry out ribbon arguments on the associated Grauert tubes over $\Cal H$, we need to lift the one form $\omega_X$, defined in Section 3.4, to its holomorphic correspondent $^{(c)}\omega_X$,
$$\align
^{(c)}\omega_X = &-\frac{1}{2}\,\,^{(c)}\overset{o}\to\varphi_{,t} (z^1,\ldots, z^n)(dx^n + idy^n) \tag6.6 \\
& - \frac{1}{2}\,\,^{(c)}\overset{o}\to\beta_{a,t} (z^1,\ldots, z^n)(dx^a + idy^a) 
\endalign
$$
with
$$\align
^{(c)}\overset{o}\to\varphi_{,t} (x^1,\ldots, x^n) & = \overset{o}\to\varphi_{,t} (x^1,\ldots, x^n) \tag6.7 \\
^{(c)}\overset{o}\to\beta_{a,t} (x^1,\ldots, x^n) &= \overset{o}\to\beta_{a,t} (x^1,\ldots, x^n),
\endalign
$$
defined on a suitable $T^s\Cal H$, where the components $^{(c)}\overset{o}\to\varphi_{,t} (z^1,\ldots, z^n)$ and $^{(c)} \overset{o}\to\beta_{a,t} (z^1,\ldots, z^n)$ each satisfy the Cauchy-Riemann equations (ensuring their holomorphicity)
$$\align
&\frac{\partial}{\partial\overline{z}^{k}}\,\, ^{(c)}\overset{o}\to\varphi_{,t} (z^1, \ldots, z^n) \tag6.8 \\
&= \frac{1}{2} (\frac{\partial}{\partial x^{k}} + i\frac{\partial}{\partial y^{k}})^{(c)}\varphi_{,t} (x^1,\ldots, x^n, y^1,\ldots, y^n) \\
&= 0\qquad k = 1, \ldots, n
\endalign
$$
and similarly for $\frac{\partial}{\partial\overline{z}^{k}} ^{(c)}\overset{o}\to\beta_{a.t}(z_1,\ldots, z^n)$.  As a holomorphic one-form $^{(c)}\omega_X$ has exterior derivative
$$\align
d^{(c)}\omega_X &= - \frac{1}{2} [\frac{\partial}{\partial z^{a}}\,\, ^{(c)}\overset{o}\to\varphi_{,t} - \frac{\partial^{(c)}}{\partial z^{n}}\overset{o}\to\beta_{a,t}] \tag6.9 \\
&\cdot (dx^a + idy^a)\wedge (dx^n + idy^n) \\
&- \frac{1}{2} \frac{\partial ^{(c)}\overset{o}\to\beta_{a,t}}{\partial z^{b}}  (dx^b + idy^b)\wedge (dx^a+ idy^a)
\endalign
$$
which, in view of the complexified Einstein equation (c.f., Equation (3.2) of Reference \cite{9}),
$$
\frac{\partial^{(c)}\overset{o}\to\varphi_{,t}}{\partial z^{a}(z)} - \frac{\partial^{(c)}\beta_{a,t}(z)}{\partial z^{n}} = 0, \tag6.10
$$
reduces to
$$
d^{(c)} \omega_X = -\frac{1}{2} \frac{\partial^{(c)}\overset{o}\to\beta_{a,t}}{\partial z^{b}} dz^b \wedge dz^a . \tag6.11
$$

For our purposes, it is convenient to regard Equation (6.11) as an equation for an ordinary, complex-valued, one form defined on a real analytic manifold of $2n$ dimensions and local coordinates
$$
\{ w^\mu \mid \mu = 1, \ldots, 2n\} = \{ x^1, \ldots, x^n, y^1, \ldots , y^n\}\tag6.12
$$
and with $^{(c)}\omega_X$ decomposed into its real and imaginary parts as 
$$
^{(c)}\omega_X = \{ (^{(c)}\omega_X^{(r)}(w))_\mu + i(^{(c)}\omega^{(i)}_X (w))_\mu\} dw^\mu . \tag6.13
$$
By appealing to the Cauchy-Riemann equations satisfied by the components, it is easy to show that the left hand side of Equation (6.11) is equal to the `ordinary' exterior derivative of $^{(c)}\omega_X$, as rewritten above, with respect to its $2n$ real coordinates $\{ w^\mu\} = \{ x^1,\ldots, x^n, y^1, \ldots, y^n\}$.  The right hand side of this equation can of course be expressed in the analogous way --- as a complex-valued two-form in the same real variables.

We are now in a position to apply Stokes's theorem much as in the previous section, the only real difference being that now the one-form in question, $^{(c)}\omega_X$ is complex and its domain of definition is a $2n$-real-dimensional Grauert tube defined over $\Cal H$.  We shall want to compare integrals of $^{(c)}\omega_X$ over different curves of the type
(6.4) extending from some `initial' slice having $x^n = \text{constant}$ to another such `final' slice.  For convenience, let us always take one such curve (which will provide a reference ``edge'' for our comparison ribbon) to lie in the real section (i.e., to have $y^a(\lambda) = y^n(\lambda) = 0$) and choose normal coordinates for $(\Sigma,\mu)$ so that points on this reference curve have \newline $x^a(\lambda) = 0$.  As in the previous section, we restrict the domain of definition of these normal coordinates to a geodesic ball relative to the metric $\mu$.  Let $p$ be the starting point of this curve so that, in the chosen coordinates $\{ x^a(p) = y^a(p) = y^n(p) = 0, x^n(p) = \overset{o}\to x^n\}$.

Now suppose that $q\in T^s\Cal H$ is a point lying in the domain of the corresponding (complex) chart and having $x^n(q) = \overset{o}\to x^n, y^n(q) = 0$, $\mu_{ab} (x^1(q),\ldots, x^{n-1}(q))y^a(q)y^b(q) < s$ where $\{ x^1(q),\ldots, x^{n-1}(q)\}$ represents a point in the aforementioned geodesic ball centered at $p$.  We want a canonical way of connecting $q$ to $p$ within the initial slice $x^n = \overset{o}\to x^n$ and, for this purpose, first connect $q$ to its projection in the real section with the `straight line'
$$\align
&x^i(\sigma) = x^i(q) = \text{constant} \\
&y^a(\sigma) = -\sigma y^a(q), \sigma\in [-1,0] \tag6.14 \\
&y^n(\sigma) = 0.
\endalign
$$
We complete the connection to $p$ along the geodesic
$$\align
&x^a(\sigma) = (1 - \sigma ) x^a(q), \,\, \sigma\in [0,1] \\
&x^n(\sigma) = x^n (p) = x^n(q) = \overset{o}\to x^n \tag6.15 \\
&y^a(\sigma) = y^n(\sigma) = 0.
\endalign
$$
This broken curve provides the starting end  (at $x^n = \overset{o}\to x^n$) for our comparison ribbon.  We complete the specification of such a ribbon by letting each point on the starting end defined above, flow along the corresponding curve of the form (6.4) (i.e., holding $x^a$ and $y^a$ constant, $y^n = 0$ and letting $x^n = \overset{o}\to x^n + \lambda$ vary until the final slice is reached).  It is easy to see, from the special form of the right hand side of 
Equation (6.11) that the corresponding two-form pulled back to such a ribbon vanishes identically and thus that Stokes's theorem applies to integrals of $^{(c)}\omega_X$ over its edges and ends in essentially the same way that we discussed in Section V for ribbons confined to the real section.  In other words, the integral of $^{(c)}\omega_X$ over the edge beginning at $q$, differs from that over the reference edge beginning at $p$ only by the (difference of) the integrals over the ribbon ends lying in the `initial' and `final' slices.

For our purposes, the contribution from the starting end, connecting $q$ and $p$, will be fixed whereas the contribution from the `final' end (connecting the images of $q$ and $p$ induced on the final slice) will vary continuously but only over a compact set (determined by the endpoint of the edge through $q$ which necessarily lies in $\overline{T^{s}\Sigma\times\bold S^{1}}$).  Thus, if as before, we designate the edges through $p$ and $q$ by $\gamma$ and $\gamma'$ respectively and the initial and final ribbon ends by $\sigma$ and $\sigma'$ respectively, then we obtain, as in the real setting,
$$\align
&\int\limits_{\gamma'}\,\, ^{(c)}\omega_X = \int\limits_\gamma\,\,^{(c)}\omega_X - (\int\limits_\sigma \,\,^{(c)}\omega_X - \int\limits_{\sigma'} \,\,^{(c)}\omega_X) \tag6.16 \\
&\qquad = \int\limits_\gamma \,\,  ^{(c)}\omega_X + \,\,^{(c)}\delta_{p,q} (x^n)  \\
\endalign
$$
with
$$
\mid^{(c)}\delta_{p,q} (\rho) \mid \leq b < \infty \quad \forall \, \rho \in [\overset{o}\to x^n, \infty). \tag6.17
$$
The integrals of course are now in general complex in value but, given the bound above, we are in a position to apply ribbon arguments to the complex setting in complete parallel to those we gave in the real setting at the end of the last section.  The arguments needed are so similar to those given previously that we shall only sketch their highlights below.

For any $q$ within the domain characterized above, we define a sequence
$$\align
&^{(c)}u_i (\overset{o}\to x^n, z^a(q)) \tag6.18 \\
&= \frac{k}{2} \int\limits_{\overset{o}\to x^{n}}^{\overset{o}\to x^n + is^*} d\rho \,\, \text{exp} [-\int\limits^\rho_{\overset{o}\to x^{n}} d\xi \frac{\overset{o}\to\varphi_{,t}}{2} (\xi, z^a(q))]
\endalign
$$
of holomorphic extensions (to $T^s\Sigma\times\bold S^1$) of the approximations given earlier in Equation (5.9) for the normalizing function $u$.  Using ribbon arguments to compare the integrals $\int_{\gamma'}\,\, ^{(c)}\omega_X$ with those for the reference curves $\int_\gamma \,\, ^{(c)}\omega_X$ we derive, as before, a bound of the form
$$\align
&\mid\,\, ^{(c)}u_m(\overset{o}\to x^n,  z^a(q)) - \,\,^{(c)}u_\ell (\overset{o}\to x^n, z^a(q))\mid \\
&\leq e^b\mid \,\,^{(c)} u_m (\overset{o}\to x^n, z^a(p)) - \,\,^{(c)} u_\ell (\overset{o}\to x^n, z^a(p))\mid \tag6.19 \\
&= e^b \mid u_m (\overset{o}\to x^n, x^a(p)) - u_\ell (\overset{o}\to x^n, x^a(p))\mid \\
&\qquad \forall \,\, \ell, m\geq 0,
\endalign
$$
where, in the final equality, we have exploited the fact that $^{(c)} u_m(\overset{o}\to x^n, z^a(p)) = u_m (\overset{o}\to x^n, x^a(p))$ by virtue of our choice that the point $p$ always lies in the real section.

As before, it follows immediately that for any $\varepsilon > 0$ there exists an $N > 0$ such that
$$
\mid\,\,^{(c)}u_m(\overset{o}\to x^n, z^a(q)) -\,\, ^{(c)}u_\ell (\overset{o}\to x^n, z^a (q))\mid < \varepsilon \quad \forall \,\, m, \ell > N
$$
and thus that the sequence $\{^{(c)}u_m(\overset{o}\to x^n, z^a(q)) \mid m = 1, 2, \ldots \}$
is Cauchy with respect to the $C^0$ norm.  Thus the sequence of approximations converges to a holomorphic limit on the domain indicated.  Repeating this argument for a (finite) collection of such domains sufficient to cover $\overline{T^s\Sigma}$ we conclude that
$$
^{(c)}u(\overset{o}\to x^n, z^a) = \frac{k}{2} \int^\infty_{\overset{o}\to x^{n}} d\rho \,\, \text{exp}[-\int^\rho_{\overset{o}\to x^{n}} d\xi \frac{\overset{o}\to\varphi_{,t}}{2}(\xi, z^a)] \tag6.20
$$
is a well-defined holomorphic function on $T^s\Sigma$ (which extends continuously to its boundary) and that, by construction, this function reduces to the real-valued function $u(\overset{o}\to x^n, x^a)$ defined in the previous section.  The latter is therefore necessarily a real-valued analytic function on $\Sigma$ which is the result we were required to prove.

\newpage

\head VII. Existence of a Killing Symmetry\endhead
\vskip .10in

We have shown that there exists a non-vanishing, analytic vector
field $K$ on $\Cal H$, tangent to the null generators of $\Cal H$ such that, in
any gaussian null coordinate chart adapted to $K$ (i.e., for which
$K$ has the local expression $K = \frac{\partial}{\partial
x^{n}}\mid_{t=0}$), the metric functions $\{\varphi ,\beta_a,
\mu_{ab}\}$ of that chart obey
$$\align
&\overset{\circ}\to\varphi = \overset{\circ}\to\beta_a =
\overset{\circ}\to\mu_{ab,n} = 0, \tag7.1 \\
&\overset{\circ}\to\varphi_{,t} = k = ~\text{constant}~\ne 0, 
\\
&(\overset{\circ}\to\beta_{a,t})_{,n} = 0 .
\endalign
$$
We shall show momentarily that $(\overset{\circ}\to\mu_{ab,t})_{,n}$
also vanishes and thus that all the metric functions and their first
t-derivatives are independent of $x^n$ on the initial surface
$t=0$ (signified as before by an overhead ``nought'').  In the
following, we shall prove inductively that all the higher time
derivatives of the metric functions are independent of $x^n$ at $t=0$
and thus that the corresponding \it analytic, \rm Lorentzian metric,
$$\align
g &= dt\otimes dx^n + dx^n\otimes dt + \varphi dx^n\otimes dx^n + \beta_a dx^a\otimes dx^n \tag7.2 \\  
&\qquad\qquad + \beta_a dx^n\otimes dx^a + \mu_{ab} dx^a\otimes dx^b,
\endalign
$$
has $\frac{\partial}{\partial x^{n}}$ as a (locally defined) Killing
field throughout the gaussian null coordinate chart considered.
Finally, we shall show that the collection of locally defined Killing
fields, obtained by covering a neighborhood of $\Cal H$ by adapted
gaussian null (agn) coordinate charts and applying the construction
mentioned above, fit together naturally to yield a spacetime Killing
field $Y$ which is analytic and globally defined on a full
neighborhood of $\Cal H$ and which, when restricted to $\Cal H$, coincides with
the vector field $K$.

Some of the results to be derived are purely local consequences of
Einstein's equations expressed in an agn coordinate chart (such as,
e.g., the observation that $\overset{\circ}\to\varphi_{,t} = k$
implies $(\overset{\circ}\to\beta_{a,t})_{,n} = 0$).  Others,
however, require a more global argument and thus demand that we
consider the transformations between overlapping, agn charts which
cover a neighborhood of $\Cal H$ in $^{(n+1)}V$.  For example, by
considering the Einstein equations $R_{ab} = 0$ restricted to $t=0$
and reduced through the use of $\overset{\circ}\to\varphi_{,t} = k =
~\text{constant}$, $\overset{\circ}\to\mu_{ab,n} = 0$ and
$(\overset{\circ}\to\beta_{a,t})_{,n} = 0$, one can derive (as in the
derivation of Equation (3.26) of Reference \cite{9}) the local equation for
$\overset{\circ}\to\mu_{ab,t}$ given by
$$
0 = -(\overset{\circ}\to\mu_{ab,t})_{,nn} + \frac{k}{2}
(\overset{\circ}\to\mu_{ab,t})_{,n}. \tag7.3
$$
Roughly speaking, we want to integrate this equation along the null
generators of $\Cal H$ and show, as in Reference \cite{9}, that it implies that
$(\overset{\circ}\to\mu_{ab,t})_{,n} = 0$.  Now, however, since the
null generators are no longer assumed to be closed curves, this
argument requires a more invariant treatment than was necessary
in Reference \cite{9}.

First, let $\{x^\mu\} = \{ t, x^n, x^a\}$ and $\{x^{\mu'}\} = \{ t',
x^{n'}, x^{a'}\}$ be any two gaussian null coordinate charts which are
adapted to $K$ (i.e., for which $K = \frac{\partial}{\partial
x^{n}}\mid_{t=0}$ and $K = \frac{\partial}{\partial
x^{n'}}\mid_{t'=0}$ on the appropriate domains of definition of the
given charts).  It is not difficult to see that, if the two charts
overlap on some region of $\Cal H$, then within that region the coordinates
must be related by transformations of the form
$$\align
x^{n'} &= x^n + h(x^a) \tag7.4 \\
x^{a'} &= x^{a'}(x^b)
\endalign
$$
where $t=t'=0$ since we have restricted the charts to $\Cal H$.  Here $h$
is an analytic function of the coordinates $\{ x^a\}$ labeling the
null generators of $\Cal H$ and $x^{a'}(x^b)$ is a local analytic
diffeomorphism allowing relabeling of those generators within the
region of overlap of the charts.

We let $\{ \varphi, \beta_a, \mu_{ab}\}$ designate the agn metric
functions of the unprimed chart,
$$\align
g &= g_{\mu\nu} dx^{\mu}\otimes dx^\nu \tag7.5 \\
&= dt\otimes dx^n + dx^n\otimes dt + \varphi dx^n\otimes dx^n + \beta_a dx^a\otimes dx^n \\
&\qquad\qquad + \beta_{a} dx^n\otimes dx^a + \mu_{ab} dx^a \otimes dx^b,
\endalign
$$
and $\{ \varphi', \beta'_{a'}, \mu'_{a'b'}\}$ designate the corresponding
functions in the primed chart.

In the region of $^{(n+1)}V$ in which the charts overlap, we have of
course,
$$
g_{\mu'\nu'} = \frac{\partial x^{\alpha}}{\partial x^{\mu'}}
\frac{\partial x^{\beta}}{\partial x^{\nu'}}  g_{\alpha\beta} \tag7.6
$$
and, because of the gaussian null metric form,
$$\align
g_{t't'} &= 0 = \frac{\partial x^{\alpha}}{\partial t'}\frac{\partial
x^{\beta}}{\partial t'} g_{\alpha\beta} \tag7.7 \\
g_{t'n'} &= 1 = \frac{\partial x^{\alpha}}{\partial t'}
\frac{\partial x^{\beta}}{\partial x^{n'}} g_{\alpha\beta} \\
g_{t'a'} &= 0 = \frac{\partial x^{\alpha}}{\partial t'}
\frac{\partial x^{\beta}}{\partial x^{a'}} g_{\alpha\beta} 
\endalign
$$
By virtue of the form of (7.4), we also have, of course, that
$\frac{\partial}{\partial x^{n}}\mid_{t=0} = \frac{\partial}{\partial
x^{n'}}\mid_{t'=0}$ on the region of overlap (since both charts were
adapted to $K$ by assumption).

Writing out Equation (7.7) in more detail, using the explicit form of
$g_{\alpha\beta}$, restricting the result to the surface $t'=t=0$ and
making use of the transformations (7.4) which hold on that surface,
one readily derives that
$$\align
&(\frac{\partial t}{\partial t'})\biggm|_{t'=0} = 1, \tag7.8 \\
&(\frac{\partial x^{a}}{\partial t'})\biggm|_{t'=0} =
(\mu^{ab}h_{,b})\biggm|_{t=0}\\ 
&(\frac{\partial x^{n}}{\partial t'})\biggm|_{t'=0} = (-\frac{1}{2}
\mu^{ab} h_{,a}h_{,b})\biggm|_{t=0}.
\endalign
$$
Differentiating these equations with respect to $x^{n'}$ and using
the fact that $\overset{\circ}\to\mu_{ab,n} = 0$ one finds that
$$
(\frac{\partial^{2}x^{\alpha}}{\partial x^{n'}\partial
t'})\biggm|_{t'=0} = 0. \tag7.9
$$
The remaining metric transformation equations (7.7), restricted to the
initial surface, yield the covariance relation
$$
\mu'_{a'b'}\biggm|_{t'=0} = (\frac{\partial x^{c}}{\partial x^{a'}}
\frac{\partial x^{d}}{\partial x^{b'}} \mu_{cd})\biggm|_{t=0} \tag7.10
$$
as well as reproducing equations such as $\varphi'\bigm|_{t'=0} = 0$,
and $\beta'_{a'}\bigm|_{t'=0} = 0$ which are common to all gaussian
null coordinate systems.

Now take the first $t'$ derivative of the transformation Equations (7.6), 
restrict the results to the surface $t' = t = 0$ and make use of
Equations (7.1) to derive expressions for
$$
\{ \varphi'_{,t'}, \beta'_{a',t'}, \mu'_{a'b',t'}\}\biggm|_{t' = 0 }
$$
in terms of unprimed quantities. Differentiating the resulting
equations with respect to $x^{n'}$ leads to the covariance relation
$$
\mu'_{a'b',t'x^{n'}}\biggm|_{t'=0} = (\frac{\partial x^{c}}{\partial
x^{a'}} \frac{\partial x^{d}}{\partial x^{b'}}
\mu_{cd,tn})\biggm|_{t=0}  \tag7.11
$$
as well as reproducing known results such as
$\beta'_{a',t'n'}\bigm|_{t'=0} = 0$ which hold in all agn coordinate
systems. 

Now in any agn coordinate chart restricted to $\Cal H$, we have the
locally defined analytic functions
$$
D \equiv \overset{\circ}\to\mu^{ac} \overset{\circ}\to\mu^{bd}
\overset{\circ}\to h_{ab} \overset{\circ}\to h_{cd} \tag7.12
$$
where $\overset{\circ}\to h_{ab}\equiv \overset{\circ}\to\mu_{ab,tn}$.
From the covariance relations (7.10) and (7.11), however, it follows that $D$ 
transforms as a scalar field in passing from one agn chart to another
in the initial surface $\Cal H$ (i.e., that $D = D'$ in the
regions of overlap).  Thus $D$ may be regarded as a globally
defined analytic function on $\Cal H$.  From the Einstein equations
$R_{ab} = 0$, restricted to $\Cal H$ and reduced by means of
$\overset{\circ}\to\varphi_{,t} = k, \overset{\circ}\to\mu_{ab,n} =
0$ and $\overset{\circ}\to\beta_{a,tn} = 0$, one can derive Equation
(7.3) in any agn chart, which in turn implies the following differential
equations for $D$:
$$
D_{,n} = kD.\tag7.13
$$
The latter can be written more invariantly as $\Cal L_KD = kD$ where
 $\Cal L_K$ represents Lie differentiation along the vector field $K$.

Equation (7.13) shows (since $k\ne 0$) that $D$ grows
exponentially along the integral curves of $K$ in $\Cal H$.  However, the
Poincar\'e  recurrence argument of Section 3.3 has shown that each
integral curve $\gamma$ of $K$, when followed arbitrarily far in
either direction from any point $p$ on $\gamma$, reapproaches $p$
arbitrarily closely.  Since $D$ is globally analytic (hence
continuous) on $\Cal H$, its values, when followed along $\gamma$, would
have to reapproach arbitrarily closely its value at $p$.  But this
is clearly incompatible with their exponential growth along $\gamma$.
The only way to avoid this contradiction arises if $D$ vanishes globally on $\Cal H$.
We thus conclude that $D = 0$ on $\Cal H$
and therefore, from the defining equation (7.12) and the fact that
$\overset{\circ}\to\mu_{ab}$ is positive definite, that
$$
\overset{\circ}\to h_{ab} = \overset{\circ}\to\mu_{ab,tn} = 0 \tag7.14
$$
on $\Cal H$.

Now, computing the first $t'$ derivatives of Equations (7.7), restricting
the results to the initial surface $t = t' = 0$ and differentiating
the resulting equations with respect to $x^{n'}$ one finds, upon
making use of Equations (7.1), (7.9) and (7.14), that
$$
\frac{\partial^{3}x^{\alpha}}{\partial x^{n'}\partial t'\partial
t'}\biggm|_{t'=0} = 0 \tag7.15
$$
whereas Equations (7.1), (7.2) and (7.14) show that
$$
(g_{\alpha\beta,tn})\biggm|_{t=0} = 0. \tag7.16
$$

We now proceed inductively to extend the above results to the case of
time derivatives of arbitrarily high order.  As an inductive
hypothesis, suppose that, for some $\ell\geq 1$ and for all $k$ such
that $0\leq k\leq\ell$, we have
$$\align
&\left(\frac{\partial}{\partial
x^{n}}\left(\frac{\partial^{k}g_{\alpha\beta}}{\partial
t^{k}}\right)\right)\biggm|_{t=0} = 0,  \tag7.17 \\
&\left(\frac{\partial}{\partial
x^{n'}}\left(\frac{\partial^{k+1}x^{\alpha}}{\partial t^{'k+1}}\right)\right) 
\biggm|_{t'=0} = 0, 
\endalign
$$
and recall that we also have
$$
\frac{\partial t}{\partial x^{n'}}\biggm|_{t'=0} = \frac{\partial
x^{a}}{\partial x^{n'}}\biggm|_{t'=0} = 0,\quad \frac{\partial
x^{n}}{\partial x^{n'}}\biggm|_{t'=0} = 1. \tag7.18
$$
Our aim is to prove that
$$\align
&(\frac{\partial}{\partial
x^{n}}(\frac{\partial^{\ell +1}g_{\alpha\beta}}{\partial 
t^{\ell +1}}))\biggm|_{t=0} = 0, \tag7.19 \\
&(\frac{\partial}{\partial
x^{n'}}(\frac{\partial^{\ell +2}x^{\alpha}}{\partial
t^{'\ell +2}}))\biggm|_{t'=0} = 0.
\endalign
$$

Note that the above imply that
$$
(\frac{\partial}{\partial
x^{n}}(\frac{\partial^{k}g_{\alpha\beta}}{\partial
x^{\gamma_{1}}\partial x^{\gamma_{2}}\ldots\partial
x^{\gamma_{k}}}))\biggm|_{t=0} = 0 \tag7.20
$$
for all $0\leq k\leq\ell$ and for arbitrary $\gamma_1, \gamma_2, \ldots
,\gamma_k$.  Furthermore, note that of the quantities
$(\frac{\partial}{\partial
x^{n}}(\frac{\partial^{\ell +1}g_{\alpha\beta}}{\partial 
x^{\gamma_{1}}\ldots\partial x^{\gamma_{\ell +1}}}))\biggm|_{t=0}$, only
$(\frac{\partial}{\partial
x^{n}}(\frac{\partial^{\ell +1}g_{\alpha\beta}}{\partial 
t^{\ell +1}}))\biggm|_{t=0}$, may be non-zero.  Now differentiate the
Einstein equation $R_{tn} = 0, \ell -1$ times with respect to $t$ and set
$t=0$ to derive an expression for $(\frac{\partial^{\ell +1}\varphi}{\partial
t^{\ell +1}})\bigm|_{t=0}$ in terms of $x^n$-invariant quantities.
Differentiate the equation $R_{tb} = 0$, $\ell -1$ times with respect to
$t$ and set $t=0$ to derive an expression for 
$(\frac{\partial^{\ell +1}}{\partial t^{\ell +1}}\beta_b)\bigm|_{t=0}$
in terms of $x^n$-invariant quantities.  Next, differentiate the
equation $R_{ab} = 0$, $\ell$ times with respect to $t$, set $t=0$ and
use the above results for 
$(\frac{\partial^{\ell +1}}{\partial t^{\ell +1}}\varphi)\bigm|_{t=0}$ and
$(\frac{\partial^{\ell +1}}{\partial t^{\ell +1}}\beta_b)\bigm|_{t=0}$,
together with those given in Equations (7.1) and (7.14) to derive an
equation of the form
$$\align
0 &= (\frac{\partial^{\ell}}{\partial t^{\ell}} R_{ab})\biggm|_{t=0} 
\\
&= - \frac{\partial}{\partial x^{n}}(\frac{\partial^{\ell +1}}{\partial
t^{\ell +1}}\mu_{ab})\biggm|_{t=0} \tag7.21 \\
&+ \pmatrix \text{positive} \\ \text{constant} \endpmatrix
\frac{\overset{\circ}\to\varphi_{,t}}{2}
(\frac{\partial^{\ell +1}}{\partial t^{\ell +1}}\mu_{ab})\biggm|_{t=0} \\
&+ \{\text{terms independent of $x^n$}\}.
\endalign
$$
Differentiate this equation with respect to $x^n$ to thus derive
$$\align
0 &= - \left(\frac{\partial^{\ell +1}\mu_{ab}}{\partial
t^{\ell +1}}\biggm|_{t=0}\right)_{,nn} \tag7.22 \\
&+ \pmatrix \text{positive} \\ \text{constant} \endpmatrix
\frac{k}{2} \left(\frac{\partial^{\ell +1}}{\partial t^{\ell +1}}
\mu_{ab}\biggm|_{t=0}\right)_{,n} 
\endalign
$$
which holds in an arbitrary agn coordinate chart.

Now define
$$\align
D^{(\ell +1)} &\equiv \frac{\det(\overset{\circ}\to h^{(\ell +1)}_{ab})}{\det
(\overset{\circ}\to\mu_{ab})} \tag7.23 \\
T^{(\ell +1)} &\equiv \overset{\circ}\to\mu^{ab} \overset{\circ}\to
h^{(\ell +1)}_{ab} 
\endalign
$$
where $\overset{\circ}\to h^{(\ell +1)}_{ab} \equiv
\left(\frac{\partial}{\partial
x^{n}}\left(\frac{\partial^{\ell +1}\mu_{ab}}{\partial
t^{\ell +1}}\right)\right)\bigm|_{t=0}$  so that Equation (7.22) becomes
$$
0 = - \overset{\circ}\to h_{ab,n}^{(\ell +1)} + \pmatrix
\text{positive}\\ \text{constant} \endpmatrix \frac{k}{2}
\overset{\circ}\to h^{(\ell +1)}_{ab} \tag7.24
$$
and $D^{(\ell +1)}$ and $T^{(\ell +1)}$ satisfy
$$\align
D^{(\ell +1)}_{,n} &= \pmatrix \text{positive} \\ \text{constant}
\endpmatrix kD^{(\ell +1)} \tag7.25 \\
T^{(\ell +1)}_{,n} &= \pmatrix \text{positive} \\ \text{constant}
\endpmatrix \frac{k}{2} T^{(\ell +1)}
\endalign
$$
in any agn coordinate chart.  To extend the Poincar\'e recurrence
argument to the quantities $D^{(\ell +1)}$ and $T^{(\ell +1)}$ we must first
show that they are globally defined analytic functions on $\Cal H$.

Differentiate the transformation equation
$$
g_{a'b'}\equiv \mu'_{a'b'} = \frac{\partial x^{\alpha}}{\partial
x^{a'}} \frac{\partial x^{\beta}}{\partial x^{b'}} g_{\alpha\beta} \tag7.26
$$
$\ell +1$ times with respect to $t'$, set $t' = 0$ and differentiate the
result with respect to $x^{n'}$.  Use the inductive hypothesis and the
vanishing of  $\left(\frac{\partial}{\partial
x^{n}}\frac{\partial^{\ell +1}\varphi}{\partial 
t^{\ell +1}}\right)\bigm|_{t=0}$  and
$\left(\frac{\partial}{\partial
x^{n}}\frac{\partial^{\ell +1}\beta_{a}}{\partial 
t^{\ell +1}}\right)\bigm|_{t=0}$ to show that this calculation yields the
covariance relation
$$\align
&\left(\frac{\partial}{\partial
x^{n'}}\frac{\partial^{\ell +1}\mu'_{a'b'}}{\partial 
t^{' \ell +1}}\right)\biggm|_{t'=0}  \tag7.27 \\
&=\left\{ \frac{\partial x^{c}}{\partial
x^{a'}}\frac{\partial x^{d}}{\partial x^{b'}}
\left(\frac{\partial}{\partial x^{n}}\frac{\partial^{\ell +1}\mu_{cd}}
{\partial t^{\ell +1}}\right)\right\}\biggm|_{t=0}
\endalign
$$
From this and Equation (7.10) it follows that $D^{(\ell +1)}$ and $T^{(\ell +1)}$
transform as scalar fields in the overlap of agn charts in $\Cal H$ and
thus that these quantities are globally defined analytic functions on
$\Cal H$.  Equations (7.25) can thus be reexpressed in the invariant form
$$\align
\Cal L_KD^{(\ell +1)} &= \pmatrix \text{positive}\\
\text{constant}\endpmatrix k D^{(\ell +1)} \tag7.28 \\
\Cal L_K T^{(\ell +1)} &= \pmatrix \text{positive} \\
\text{constant}\endpmatrix \frac{k}{2} T^{(\ell +1)}
\endalign
$$
and show that $D^{(\ell +1)}$ and $T^{(\ell +1)}$ grow exponentially (unless
they vanish) when followed along the integral curves of $K$ in $\Cal H$
(i.e., along the null generators of $\Cal H$).  Repeating the Poincar\'e
recurrence argument given previously for $D$ now yields a
contradiction unless $D^{(\ell +1)}$ and $T^{(\ell +1)}$ vanish globally on
$\Cal H$.  This in turn implies that
$$
\left(\frac{\partial}{\partial
x^{n}}\frac{\partial^{\ell +1}\mu_{ab}}{\partial
t^{\ell +1}}\right)\biggm|_{t=0} = 0 \tag7.29
$$
in every agn chart on $\Cal H$ and, together with the results obtained
above for the other metric components, shows that 
$$
\left(\frac{\partial}{\partial
x^{n}}\frac{\partial^{\ell +1}g_{\alpha\beta}}{\partial
t^{\ell +1}}\right)\biggm|_{t=0} = 0 \tag7.30
$$
in every such chart.

Applying the technique of the previous paragraph to the
transformation equations for $\varphi'$ and $\beta'_{a'}$ merely
produces covariance relations for the quantities
\newline $\left(\frac{\partial}{\partial
x^{n}}\left(\frac{\partial^{\ell +1}\varphi}{\partial
t^{\ell +1}}\right)\right)\bigm|_{t=0}$ and
$\left(\frac{\partial}{\partial
x^{n}}\left(\frac{\partial^{\ell +1}\beta_{a}}{\partial
t^{\ell +1}}\right)\right)\bigm|_{t=0}$ which are consistent with the (already
established) vanishing of these quantities in every agn chart.  To
complete the inductive proof, we differentiate the remaining
transformation equations (7.7) $\ell +1$ times with respect to $t'$, set
$t'=0$, use the inductive hypothesis and the new results summarized
in Equation (7.30) to show that
$$
\left(\frac{\partial}{\partial
x^{n'}}\frac{\partial^{\ell +2}x^{\alpha}}{\partial
t^{'\ell +2}}\right)\biggm|_{t'=0} = 0. \tag7.31
$$
This result, together with that of Equation (7.30), completes the proof by
induction. 

It follows from the analyticity of $g$ and the inductive proof given
above that $\left(\frac{\partial}{\partial x^{n}}
g_{\alpha\beta}\right)$ vanishes throughout any agn coordinate chart
and thus that $K \equiv\frac{\partial}{\partial x^{n}}$ is a (locally
defined) analytic Killing field throughout the given chart. In the
region of overlap of any two such charts we have the two locally
defined Killing fields $K = \frac{\partial}{\partial x^{n}}$ and $K'
= \frac{\partial}{\partial x^{n'}}$ and we wish to show that, in
fact, they coincide. By construction both $K$ and $K'$ coincide 
on their appropriate domains of definition within the null
surface $\Cal H$. Therefore $Z\equiv K'-K$ is an analytic Killing field
of $g$ defined locally on the region of overlap of the two charts
which vanishes on the intersection of this region with the null
surface $\Cal H$. This implies that $Z$ vanishes throughout its domain of
definition, however, since the Killing equations
$$
Z_{\mu,t} + Z_{t,\mu} - 2 ~~^{(n+1)}\Gamma^\nu_{\mu t}Z_\nu = 0 \tag7.32
$$
determine $Z$ uniquely from data $Z\bigm|_{t=0}$ (in the analytic case)
and have only the trivial solution $Z=0$ if $Z\bigm|_{t=0} = 0$.

It follows from the above that there exists a unique analytic Killing
field $K$, globally defined on a full neighborhood of $\Cal H$ in
$(^{(n+1)}V,g)$ which, when restricted to $\Cal H$, coincides with the
vector field $K\mid_{\Cal H}$ constructed thereon and thus is tangent to the null generators of $\Cal H$.
In fact, one can prove that $K$ extends to a Killing field defined
throughout the maximal Cauchy development of the globally hyperbolic
region of $(^{(n+1)}U,g)$ whose Cauchy horizon is $\Cal H$. The techniques
for proving this were discussed at the end of Section III of Reference \cite{9}
and need not be repeated here. One can further show, by a straightforward computation, that
$$
\left\{ K^{\beta} ~~^{(n+1)}\nabla_\beta K^\alpha + \frac{k}{2}
K^\alpha\right\}\biggm|_{\Cal H} = 0.  \tag7.33
$$
To obtain the properly normalized horizon generating Killing field $Y$ we shall need to rescale $K$ with a suitably chosen multiplicative constant. This in turn will rescale $-\frac{k}{2}$ to the true surface gravity $\kappa$ of the horizon (c.f. Equation (1.1)).

First of all however, we need to show that the commutator, $[K,T]$, of the newly constructed Killing field $K$ with the stationarity generator $T$ vanishes everywhere on $\{^{(n+1)}U, g\}$.  We do this by first proving that $[K,T]\mid_{\Cal H} = 0$ and then by recalling an argument that shows that a Killing field on a connected (pseudo-) Riemannian manifold which, together with its first covariant derivative, vanishes at a point of that manifold must in fact vanish everywhere.

We began the construction of $K$ with the choice of an analytic vector field $X\mid_{\Cal H}$, defined on $\Cal H$, that was everywhere tangent to the null generators of $\Cal H$ and that commuted with $T\mid_{\Cal H}$.  We then `renormalized' $X\mid_{\Cal H}$ with an analytic function $u$, defined on $\Cal H$, constructed such that the past affine length of the incomplete null generator `starting' with initial conditions $(p, K(p)) = (p, u(p) X\mid_{\Cal H}(p))$ will have a fixed numerical value $2/k$, for any point $p\in\Cal H$.  In coordinates adapted to $T$ and a foliation of 
$\Cal H$ generated therefrom (i.e., having $T = \frac{\partial}{\partial x^n}$ and leaves defined by $x^n$ = constant) it is not difficult to see that the construction of $u$ from $X\mid_{\Cal H}$ and a spacetime metric independent of $x^n$ in the chosen coordinates, necessarily satisfies $T\mid_{\Cal H} u = \frac{\partial}{\partial x^n} u\mid_{\Cal H} = 0$ (i.e.,  that the needed renormalizations factor $u$ is invariant with respect to the flow on $\Cal H$ generated by $T\mid_{\Cal H}$.

Extending $K$ as we have done to the ambient `cosmological' spacetime $\{^{(n+1)}U,g\}$ we define the commutator Killing field $Z:= [K,T]$ thereon and conclude, from the above that at least $Z$ vanishes everywhere on the hypersurface $\Cal H$.  But from Killing's equations and the vanishing of $Z$ on $\Cal H$, it follows easily that both $Z_\mu(p)$ and \newline
$Z_{\mu;\nu}(p):= ^{(n+1)}\nabla_\nu Z_\mu (p)$ vanish at any point $p\in\Cal H$ (where $Z_\mu =  g_{\mu\nu} Z^\nu$).  However, a standard argument (c.f. Reference \cite{23}, Appendix C.3) shows that a Killing field $Z$ in any connected, pseudo-Riemannian manifold is uniquely determined, at an arbitrary point $q$, through the integration of a system of linear first order, homogeneous ordinary differential equations along an arbitrary differentiable curve leading from $p$ to $q$, in terms of the initial conditions defined by $\{Z_\mu (p), Z_{\mu;\nu}(p)\}$. When these data vanish, the solution Killing field necessarily vanishes everywhere.  We thus conclude that the commutator $[K,T]$ not only vanishes on $\{^{(n+1}U,g\}$ but also on any, connected embedding spacetime to which $K$ and $T$ can be extended by analytic continuation.

We now have two, commuting, analytic Killing fields defined on the quotient, `cosmological' spacetime $(^{(n+1}U, g)$- the stationarity generator $T$ and the Killing horizon generator $K$.  Thanks to the identifications made in constructing this quotient, $T$ has closed orbits, hence generates a circle action, whereas $K$, for the cases of most interest here, has non-closed orbits.  Before analyzing this generic situation however, let us briefly consider the special case in which $K$ also has closed orbits.

In this case, both $T$ and $K$ generate circle actions on $\Cal H$ and, upon rescaling $K$ with a suitable constant, we can define a Killing field $Y$ whose integral curves close after the same lapse of curve parameter as do those of $T$ (i.e., after the lapse $s^*$ defined in Section 3.1).  It is worth noting here that the closure of the orbits of $K$ (hence $Y$) will depend in general upon the choice of $s^*$ which otherwise is arbitrary.

In this setup, $Y-T$ is a Killing field in $(^{(n+1)}U, g)$ that generates a circle action on $\Cal H$ which takes any $x^n =$ constant slice to itself (after a `closure' parameter lapse $s^*$) whereas $Y$ and $T$ separately would map this surface to one labeled by $x^n + s^*$ (which however, is identified with the original one in defining the quotient).  The circle action generated by $Y-T$ will thus persist to yield a circle action on $\tilde\Cal H$ when we relax the aforementioned identifications, whereas those generated by $Y$ and $T$ separately will be unwrapped to $\bold R$-actions on $\tilde\Cal H$.

Thus we can now write $Y=T + \Omega\Phi$ where $\Omega$ is a constant and $\Phi$ is a Killing field with orbits which remain closed even after the aforementioned identifications are relaxed.  One normalizes the choice of $\Phi$ and $\Omega$ by requiring that the orbits of $\Phi$ close with the canonical lapse for an ``angular'' parameter, namely $2\pi$.  Note however that a further convention would be needed to fix uniquely the algebraic signs of $\Phi$ and $\Omega$ since otherwise they could be replaced by $-\Phi$ and $-\Omega$ respectively.
The Killing field $\Phi$ can be naturally interpreted as an `axisymmetry generator' and the constant $\Omega$ the corresponding `angular velocity' of the black hole in the symmetry direction.  Note however that we are {\it not} claiming that $\Phi$ must necessarily vanish somewhere and thus that an actual axis of rotational symmetry must necessarily exist.

\newpage

\head VIII. The Black Hole Isometry Group\endhead
\vskip .05in

In gaussian null coordinates adapted to the (horizon generating) Killing field $K$, we have
$$
\Theta := g_{\alpha\beta} K^\alpha K^\beta = \varphi \tag8.1
$$
and
$$\align
&g^{\mu\nu} \Theta_{,\mu} \Theta_{,\nu} = g^{\mu\nu} \varphi_{,\mu} \varphi_{, \nu}  \tag8.2 \\
&= (-\varphi + \beta_a\beta^a) \varphi_{,t}\varphi_{,t} - 2\beta^a \varphi_{,a} \varphi_{,t} \\
&\qquad + \mu^{ab} \varphi_{,a} \varphi_{,b} 
\endalign
$$
where terms involving $\varphi_{,n}$ vanish since $K = \frac{\partial}{\partial x^{n}}$ is Killing.  The null surface $\Cal H$ is of course a level surface of $\Theta$ since $\Theta\mid_{\Cal H} = \overset{\circ}\to\varphi = 0$ but, from Equation (8.2), one finds that
$$
\frac{\partial}{\partial t} (g^{\mu\nu} \Theta_{,\mu} \Theta_{,\nu})\mid_{t=0} = - (\overset{\circ}\to\varphi_{,t})^3 = - k^3 < 0 \tag8.3
$$
so that, whereas the gradient of $\Theta$ is null on $\Cal H$, it is timelike on some (sufficiently small) open
set bounded by $\Cal H$ that corresponds to $t > 0$ in the agn coordinate systems used in the calculation.  Thus the level surfaces of $\Theta$ will be spacelike hypersurfaces, diffeomorphic to $\Sigma\times\bold S^1$, that foliate an open set lying to one side of $\Cal H$ (corresponding to values of $\Theta$ in the range $(0,a)$ for sufficiently small $a > 0$). Let us designate this open set by $^{(n+1)}W$ and the corresponding, `cosmological' spacetime by $\{ ^{(n+1)}W, g\}$. By a well-known result \cite{24} $\{ ^{(n+1)}W, g\}$ will be globally hyperbolic, having the level surfaces of $\Theta$ as Cauchy hypersurfaces.  By contrast the opposite side of $\Cal H$ will be acausal, having the orbits of $K$ as nearly closed timelike curves.  The null surface $\Cal H$  serves as a Cauchy horizon separating these two cosmological regions.  In familiar examples $\{ ^{(n+1)}W, g\}$ corresponds to a region interior to the original black hole solution but we shall not assume that  this is always the case ---- perhaps in some `exotic' higher dimensional examples it could correspond to an exterior region.

From Killing's equations for $K$ and $T$ and the fact that $K$ and $T$ commute, it is straightforward to show that
$$
T^\mu \Theta_{,\mu} = K^\mu \Theta_{,\mu} = 0 \tag8.4
$$
and thus that $T$ and $K$ are both tangent to the Cauchy hypersurfaces of $\{ ^{(n+1)}W, g\}$ defined by $\Theta =$ constant.  It follows from standard results on the Killing symmetries of vacuum spacetimes that the first and second fundamental forms induced by $g$ on each such $\Theta =$ constant Cauchy surface are both invariant under the flows on these surfaces generated by the two corresponding tangential vector fields. $T$ of course generates a circle action on each such surface whereas $K$, in the generic case of most interest, generates an $\bold R$ action. Thus the action generated by the pair is generically non-compact.

But the isometry group of any compact Riemannian manifold (such as any one of these Cauchy surfaces) is necessarily a compact Lie group and since the action generated thereon by $K$ and $T$ is abelian, one can show that the full isometry group must contain a toral subgroup of dimension at least $2$ \cite{11}.  In fact the arguments of this reference (which, though framed there primarily for $3+1$ dimensions, apply in arbitrary dimensions) establish that both the first and second fundamental forms of such a Cauchy hypersurface must necessarily be simultaneously invariant under a toral action $\bold T^m$ for some $m\geq 2$. The dimension $m$ of this action is determined by the dimension of the closures of the orbits of $\{ T,K\}$ on any of the Cauchy hypersurfaces defined by $\Theta =$ constant.

The existence of a toral action which leaves both first and second fundamental forms of a Cauchy hypersurface invariant implies, by standard results on the Killing symmetries of globally hyperbolic, vacuum spacetimes \cite{25}, that there exists on $\{ ^{(n+1)}W, g\}$ a set of spacetime Killing fields $\{ \Gamma_1, \Gamma_2,\ldots, \Gamma_{m-1}, \Gamma_m = T\}$ that generate this toral action where each $\Gamma_i$ (with $\Gamma_m = T$) generates a circle action and where $[\Gamma_i,\Gamma_j] = 0$ for all $i,j\in [1,\ldots ,m]$. Since the action of $K$ must be included therein, we must have
$$
K = \alpha_1\Gamma_1 +\ldots+ \alpha_{m-1}\Gamma_{m-1} + \alpha_m T \tag8.5
$$
for some constants $\alpha_1,\ldots, \alpha_m$.

Now the circle actions generated by some of the $\Gamma_i$, for $1\leq i\leq m-1$, might possibly `unwrap' to $\bold R$-actions on the covering manifold $^{n+1}\tilde W$ when the compactification identifications are relaxed (as that of $\Gamma_m = T$ always does) but, following the construction given at the end of the previous section, one can always `shift' each $\Gamma_i, 1\leq i\leq m-1$ (if necessary) by a suitably chosen multiple of $T$ so that the circle actions generated by these shifted $\Gamma_i$'s remain compact under the unwrapping (though of course that generated by $\Gamma_m = T$ will not). Thus, without loss of generality, we can assume that each of the $\Gamma_i, 1\leq i\leq m-1$ has been chosen so that its circular orbits survive the ultimate `unwrapping' and, after a further trivial rescaling, that each one has the natural orbit period $2\pi$, appropriate for a generator of rotations.

Since the orbits of $K$ must correspondingly unwrap to yield future complete horizon generators of $\bold{\tilde\Cal H}$ on the covering spacetime it's clear that $\alpha_m$ must be non-vanishing in Equation (8.5). Dividing by $\alpha_m$, setting $\Omega_i = \frac{\alpha_{i}}{\alpha_{m}}$, $1\leq i\leq m-1$ and defining $Y = \frac{K}{\alpha_{m}}$ we get
$$
Y = T + \sum\limits^{m-1}_{i=1} \Omega_i\Phi_i \tag8.6
$$
where $\Phi_i$ designates $\Gamma_i$ rescaled to have orbit period $2\pi$. From the foregoing, we know that all of the Killing fields $\{ Y,T,\Phi_i\}$ mutually commute.

We already know that $T$ and $Y$ are analytic and extend, through $\Cal H$, to the other side of this horizon.  But to analytically extend the individual $\Phi_i$ we need, first of all, to know whether they are analytic on  $\{^{(n+1)}W,g\}$.
That any Killing field $Z$ is in fact analytic (on a neighborhood of any point $p\in ^{(n+1)}W$) on this space can be established through an application of the Cauchy-Kowalewski theorem \cite{19} to the standard system of linear differential equations that one uses to propagate a Killing field from initial data $\{ Z^\mu (p), Z^{\mu}_{;\nu} (p)\}$ given at an arbitrary point $p$ \cite{23, 26}.  In analytic charts for $g$, one parametrizes a neighborhood of $p$ by the coordinate curves
$$\align
& x^\mu (\lambda) = x^\mu (p) + v^\mu \lambda, \tag8.7 \\
& 0\leq \lambda \leq 1 ,\,\, v^\mu = x^\mu - x^\mu (p)
\endalign
$$
to represent arbitrary nearby points as the endpoints (at $\lambda =1$) of these curves.  Solving the differential system with the chosen initial conditions and setting $\lambda = 1$ yields $Z(x)$ and guarantees analyticity of the solution with respect to $\{ x^\alpha\}$ (since the coefficients of this linear system are analytic in $\{ x^\alpha, \lambda\}$). Strictly speaking the differential system in question involves only ordinary differential equations in that no partial derivatives with respect to the $\{ x^\alpha\}$ actually occur but, nevertheless, the Cauchy-Kowalewski theorem applies to this system and implies joint analyticity in the independent variables $\{ x^\alpha,\lambda\}$ which, after setting $\lambda = 1$, yields the desired result for $Z(x)$.

Now, by starting at points of $^{(n+1)}W$ one can compute an extension of each $\Phi_i$ (to the full tubular neighborhood $^{(n+1)}U$ of the horizon $\bold{\Cal H}$)  by integrating the aforementioned linear differential systems along (for example) the integral curves of the transversal null field $L = \frac{\partial}{\partial t}$ introduced in Section 3.1. The danger though is that, whereas such extensions are guaranteed to be locally analytic (by the Cauchy-Kowalewski argument now applied to arbitrary points of $^{(n+1)}U$), there is no a priori guarantee that all possible such extensions lead to a single valued (analytic) Killing field defined throughout $^{(n+1)}U$.  For example, suppose that $p$ and $q$ are two points of $^{(n+1)}U$, with $p\in ^{(n+1)}W$, and that one computes $\Phi_i(q)$ by integration along some curve $\gamma$ connecting $p$ to $q$.  If $\gamma'$ is another curve in $^{(n+1)}U$ connecting these same points, then one could equally well have computed a (possibly different) $\Phi'_i (q)$ by integrating along $\gamma'$.  If $\gamma'$ is homotopic to $\gamma$ however, then a theorem of Nomizu's ensures that this cannot happen and that $\Phi'_i(q)$ will coincide with $\Phi_i(q)$ \cite{26}.  But $^{(n+1)}U \approx \Sigma\times\bold S^1\times I$ (for some interval $I$) and thus is not simply connected by virtue of the circular factor (and perhaps also the topology of $\Sigma$).  Suppose therefore that $\gamma'$ is not homotopic to $\gamma$ -- can we still show that $\Phi'_i(q)$ necessarily coincides with $\Phi_i(q)$?  To see that we can, notice that $\gamma'$ can be homotopically deformed (by pushing its points along integral curves of $L$ for example) to a curve that lies entirely within $^{(n+1)}W$ (intercepting the original curve $\gamma$ at some intermediate point $r$) together with the segment of $\gamma$ which connects  $r$ to $q$.  But propagation of $\Phi_i$  along the new curve segment between $p$ and $r$ is locally unique and lies entirely within the manifold $^{(n+1)}W$ where $\Phi_i$ is globally defined and so must necessarily reproduce $\Phi_i(r)$.  Further propagation along the segment of $\gamma$ from $r$ to $q$ is identical to that for the original calculation and thus leads to the conclusion that $\Phi'_i(q) = \Phi_i(q)$ as desired.

It is conceivable of course that the `cosmological' spacetime $\{^{(n+1)}W,g\}$ admits other globally defined Killing fields, beyond those implied by our argument (and perhaps failing to commute with them).  If so, they could also be propagated to uniquely defined, analytic Killing fields on $\{^{(n+1)}U,g\}$ by the techniques sketched above and thereon yield generators of the full, continuous isometry group of $\{^{(n+1)}U,g\}$.  If any such additional Killing fields exist, let us designate them $\chi_1,\ldots, \chi_r$ for some $r\geq 1$.

Now, one can lift each of the Killing fields $\{ Y, \Phi_1,\ldots, \Phi_{m-1}, T, \chi_1,\ldots,\chi_r\}$ (with $Y$ given in terms of $T$ and the $\{\Phi_i\}$ by Equation (8.6)) back to the original covering manifold $\{^{(n+1)}\tilde U,\tilde g\}$ to get a corresponding set of Killing fields $\{\tilde Y, \tilde\Phi_1,\ldots, \tilde\Phi_{m-1}, \tilde T, \tilde\chi_1,\ldots, \tilde\chi_n\}$ defined on a tubular neighborhood $^{(n+1)}\tilde U$ of the original black hole horizon 
$\tilde\Cal H$. It is possible as well that this covering manifold $\{^{(n+1)}\tilde U,\tilde g\}$ admits some additional (necessarily analytic) Killing fields that are not compatible with the quotienting procedure used to define 
$^{(n+1)}U$. If so, they could be added to the list above supplementing the set of $\tilde\chi_k$'s.

We now wish to analytically extend each such Killing field to the full spacetime  $\{^{(n+1)} \tilde V,\tilde g\}$ but again the potential for a breakdown of uniqueness arises in view of the fact that $^{(n+1)}\tilde V$ is not, in general, simply connected. Fortunately, however, the {\it topological censorship} theorem \cite{27} applies to vacuum, asymptotically flat black hole spacetimes in all dimensions and asserts that the {\it domain of outer communications} is necessarily simply connected provided that null infinity, $\Cal I$, is assumed to have this property. Thus by assuming a simply connected $\Cal I$ and appealing to topological censorship, we can avoid the obstructions to applying Nomizu's theorem and uniquely, analytically extend each of the Killing fields described above to the full domain of outer communications. In fact, we can do a little more, by appealing to a straightforward generalization of the argument sketched above, and analytically extend each Killing field to a full tubular neighborhood of each horizon component, thereby enlarging its domain of unique definition to at least a portion of the region interior to the black holes. If, as in familiar examples, the spacetime also admits past horizons bounding white hole regions the arguments of this article can of course be applied, in a time reversed sense, to extend each Killing field to a portion of the region interior to the white holes as well. \newline
\indent It may seem incredible that topological censorship is not contradicted by the existence of black ring \cite{2} and black Saturn \cite{4} solutions already in five dimensions, but a key point is that the rings in these examples have $\bold S^2\times \bold S^1$ topologies. By deforming a loop that links one of these rings until it lies in an $\bold R^3$ hyperplane that intersects the horizon transversal to the $\bold S^1$ ``direction'', one sees that this loop merely passes around a copy of $\bold S^2$ and so can easily be slipped off and contracted to a point. This magician's trick would not have been possible had the horizon's cross-sectional topology been $\bold T^3$ for example, instead of $\bold S^2\times \bold S^1$ but an independent argument by Galloway and Schoen has shown that a stationary horizon's cross-sectional topology must be that of a positive Yamabe class manifold (and thus admit a metric of strictly positive scalar curvature) \cite{28}.

\newpage

\head IX. Concluding Remarks\endhead
\vskip .10in

A curious feature of the construction presented here is that it can only detect the extra Killing fields $\{ \Phi_1,\ldots, \Phi_{m-1}\}$ if, when taken together with $T = \Phi_m$, the linear combination
$$
Y = T + \sum\limits^{m-1}_{i=1} \Omega_i \Phi_i ,
$$
yielding the horizon generator $Y$, has orbits that densely fill $m$-dimensional tori.  For this to be true, none of the $\Omega_i$ can be rational multiples of $\frac{2\pi}{s^{*}}$ and each of the ratios  $\Omega_j/\Omega_{\ell}$ must also be irrational since otherwise only lower dimensional tori could then be densely filled by the orbits of $Y$.  On the other hand, it seems implausible that solutions of the field equations having a fixed set of Killing fields  $\{\Phi_1,\ldots, \Phi_{m-1}, T\}$ would only exist for incommensurable values of their associated angular velocities $\{ \Omega_1,\ldots, \Omega_{m-1}\}$. It seems more plausible that families of solutions (sharing a fixed set of Killing fields) should exist with continuously adjustable values of their corresponding angular velocities.
 But, if so, then when all of the $\Omega_i$'s are rational multiples of $\frac{2\pi}{s^{*}}$ (for a suitably chosen $s^*$) the orbits of $Y$ would be closed and the principal arguments of the present article would no longer apply.

For these (`closeable generator') cases however, the straightforward extension of our earlier arguments (to the higher dimensional settings under study here) does apply and again yields existence of the (horizon generating) Killing field $Y$ \cite{9, 10}.  In fact, these arguments even apply to the {\it degenerate cases} for which the horizon generators are complete in both directions and different techniques are needed to define the candidate Killing fields on the horizons and to propagate them into the embedding spacetimes. This was done (in the four-dimensional setting) for the Einstein-Maxwell equations in Reference \cite{9} but generalizes at once to higher dimensional problems with closeable generators.

On the other hand, when the  orbits of $Y$ are closed, one cannot use the arguments of \cite{11}, as we have done here, to deduce the existence of additional Killing fields $\{\Phi_1,\ldots, \Phi_{m-1}\}$, though of course, for stationary solutions, the presence of $T$ is assumed a priori, and one can then always define $\Phi_1$ such that
$$
Y = T + \Omega_1 \Phi_1.
$$
In the purely cosmological setting of vacuum spacetimes admitting compact, analytic Cauchy horizons (wherein the `extra' Killing field analogous to $T$ here is {\it not} assumed to exist), analytic solutions to the field equations (at least in four-dimensions but most probably in higher dimensions by a straightforward extension of the arguments used) need not admit {\it any} additional Killing fields at all.  One can use Cauchy-Kowalewski techniques to prove the existence of large families of solutions (parametrized by arbitrary, adjustable {\it functions} rather than merely adjustable constants such as the $\Omega_i$'s) that generically have no Killing fields beyond that given by the horizon generator, $Y$ \cite{29, 30}.  Thus it can be significantly misleading (at least in the cosmological context) to imagine that solutions with closed generators are simply special cases of solution families with additional Killing fields that are attained when certain adjustable parameters are tuned to have `rational' values. The closed orbit cosmological solutions are far less rigid than their non-closed orbit counterparts. For a stationary, rotating black hole spacetime with closeable generators, one will always have $\Phi_1$, since this is just a linear combination of $T$ and $Y$, but it seems possible that no additional Killing fields need exist --- the corresponding symmetries having perhaps been broken by external distributions of matter. Though in practice one often finds solutions with closeable generators comprising `dense subsets' of solution families admitting multiple axisymmetry generators (c.f. References \cite{1-6} for example), the former almost surely belong to much larger sets of `perturbed' (stationary, rotating) black hole spacetimes having, generically, only $2$-dimensional isometry groups. It would be interesting to know however, whether such `symmetry breaking perturbations' can persist for {\it isolated} stationary black holes, undisturbed by external sources of matter.

\centerline{\bf Acknowledgment}
\medskip

This article was somewhat delayed in its completion but is based primarily upon the authors' work during visits to the Isaac Newton Institute in Cambridge, England and the Kavli Institute for Theoretical Physics in Santa Barbara, California in late 2005 and early 2006.  Further work thereon was carried out with subsequent visits to the Albert Einstein Institute in Golm, Germany, the Erwin Schr\"odinger Institute in Vienna, Austria and the Institut des Hautes \'Etudes Scientifiques in Bures-sur-Yvette, France.  The authors are very grateful for the hospitality and support given to them during these visits by each of the institutes mentioned above.  The authors also benefitted significantly from interactions with Stefan Hollands, Akihiro Ishibashi, Robert Wald, Piotr Chru\'sciel and Gregory Galloway.  This research was also supported by the National Science Foundation through the grants PHY-0354659 and PHY-0652903 to the University of Oregon and the grants PHY-0354391 and PHY-0647331 to Yale University.

\newpage

\baselineskip 10pt
\centerline{\bf References}
\vskip .10in

\noindent [1] R. Myers and M. Perry, {\it Black holes in higher dimensional space-times},
Ann. Phys. 

{\bf 172}, 304-347 (1986).
\vskip .10in

\noindent [2] R. Emparan and H. Reall, {\it Black holes in higher dimensions},
gr-qc/0801.3471
\vskip .10in

\noindent [3] R. Emparan and H. Reall, {\it Black rings}, Class. Quant. Grav.
{\bf 23} R169 (2006).
\vskip .10in

\noindent [4] H. Elvang and P. Figueueras, {\it Black saturns}, hep-th/070135
\vskip .10in

\noindent [5] G. Gibbons, D. Ida and T. Shiromizu, {\it Uniqueness and non-uniqueness of static 

vacuum black holes in higher dimensions}, Prog. Theor. Phys.  Suppl. {\bf 148}, 284-290 

(2003).
\vskip .10in

\noindent [6] G. Horowitz, {\it Higher dimensional generalizations of the Kerr black hole}, in {\it Kerr 

Spacetime: Rotating Black Holes} (S. Scott, M. Visser and D. Wiltshire, eds.) 

Cambridge University Press, (2008).
\vskip .10in

\noindent [7] P. Chrusciel and R. Wald, {\it Maximal hypersurfaces in asymptotically stationary 

space-times}, Comm. Math. Phys. {\bf 163}, 561-604 (1994).
\vskip .10in

\noindent [8] G. Comp\'ere, {\it An introduction to the mechanics of black holes},
Lecture notes prepared 

for the Second Mondave Summer School in Mathematical Physics, gr-qc/0611129v1
\vskip .10in

\noindent [9] V. Moncrief and J. Isenberg, {\it Symmetries of cosmological Cauchy horizons}, Comm. 

Math. Phys. {\bf 89}, 387-413 (1983).
\vskip .10in

\noindent [10] J. Isenberg and V. Moncrief, {\it Symmetries of cosmological Cauchy horizons with

~~~ exceptional orbits}, J. Math. Phys. {\bf 26}, 1024-1027 (1985).
\vskip .10in

\noindent [11] J. Isenberg and V. Moncrief, {\it On spacetimes containing Killing vector fields with

~~~ non-closed orbits}, Class. Quant. Grav. {\bf 9}, 1683-1691 (1992).
\vskip .10in

\noindent [12] P. Chrusciel, {\it On rigidity of analytic black holes}, Comm. Math. Phys. {\bf 189},

~~~ 1-7 (1997).
\vskip .10in

\noindent [13] S. Hollands, A. Ishibashi and R. Wald, {\it A higher dimensional stationary rotating 

~~~ black hole must be axisymmetric}, Comm. Math. Phys. {\bf 271}, 699-722 (2007).
\vskip .10in

\noindent [14] The techniques for doing this were first developed by H. Friedrich, Istv\'an R\'acz 

~~~ and R. Wald in {\it On the rigidity theorem for spacetimes with a stationary event

~~~ horizon or a compact Cauchy horizon}, Commun. Math, Phys. {\bf 204}, 691-707 (1999).
\vskip .10in

\noindent [15] C. B. Morrey, {\it Multiple Integrals in the Calculus of Variations}, Springer (1966). 

~~~ See in particular Theorem 6.7.6 on p. 271.
\vskip .10in
 
\noindent [16] S. Hawking and G. Ellis, {\it The Large Scale Structure of Space-Time}, Cambridge

~~~ Univ. Press, (1973). See especially Section 8.5.
\vskip .10in

\noindent [17] H. Poincar\'e, {\it Les M\'ethodes Nouvelles de la M\'ecanique Classique C\'eleste} (Vol. 3) 

~~~ Gauthiers-Villars, (1899).
\vskip .10in

\noindent [18] V. Arnold, {\it Mathematical Methods of Classical Mechanics}, Springer (1978).
\vskip .10in

\noindent [19]  F. John, {\it Partial Differential Equations}, Springer (1991). See especially 

~~~ chapter 3.
\vskip .10in

\noindent [20] F. Treves, {\it Basic Linear Partial Differential Equations}, Dover Publications (2006). 

~~~ See in particular Proposition 17.1
\vskip .10in

\noindent [21] For a classical treatment of this issue see Theorem 1.6 in {\it Analysis of Several 

~~~ Complex Variables} by T. Ohsawa, American Mathematical Society (1998).
\vskip .10in

\noindent [22] D. Burns, S. Halverscheid and R. Hind, {\it The geometry of Grauert tubes and 

~~~ complexification of symmetric spaces}, Duke Math. J. {\bf 118}, 465-491 (2003).
\vskip .10in

\noindent [23] R. Wald, {\it General Relativity}, Univ. Chicago Press (1984).
\vskip .10in

\noindent [24] R. Budic, J. Isenberg, L. Lindblom and P. Yasskin, {\it On the determination of Cauchy 

~~~ surfaces from intrinsic properties}, Comm. Math. Phys. {\bf 61}, 87-95 (1978).
\vskip .10in

\noindent [25] A. Fischer, J. Marsden and V. Moncrief, {\it The structure of the space of solutions of 

~~~ Einstein's equations I. One Killing field}, Ann. Inst. H. Poinc. Sect. A {\bf 33}, 147-194 

~~~ (1980).
\vskip .10in

\noindent [26] K. Nomizu, {\it On local and global existence of Killing vector fields}, Ann. Math.

~~~ {\bf 72}, 105-120 (1960).
\vskip .10in

\noindent [27] G. Galloway, K. Schleich, D. Witt and E. Woolgar, {\it Topological censorship 

~~~ and higher genus black holes}, Phys. Rev. D. (3) {\bf 60} 104039 (1999).
\vskip .10in

\noindent [28] G. Galloway and R. Schoen, {\it A generalization of Hawking's black hole topology 

~~~ theorem to higher dimensions}, Comm. Math. Phys. {\bf 266}, 571-576 (2006).
\vskip .10in

\noindent [29] V. Moncrief, {\it The space of (generalized) Taub-NUT spacetimes}, J. Geom. Phys. {\bf 1},

~~~ 107-130 (1984).
\vskip .10in

\noindent [30] V. Moncrief, {\it Neighborhoods of Cauchy horizons in cosmological spacetimes with 

~~~ one Killing field}, Ann. Phys. (N.Y.) {\bf 141}, 83-103 (1982).

\end